\documentclass[pra,9pt, twocolumn, aps, superscriptaddress, showpacs, groupaddress]{revtex4-1}
\usepackage[T1]{fontenc}
\usepackage{bigints}
\usepackage{amssymb}
\usepackage{graphicx}
\usepackage{amsmath}
\usepackage{dcolumn}
\usepackage{multirow}
\usepackage{hyperref}
\usepackage[normalem]{ulem}
\usepackage{physics}
\usepackage{bm}
\usepackage{comment}
\usepackage{xcolor}
\usepackage[caption=false]{subfig}
\usepackage{siunitx}
\usepackage{dsfont}
\usepackage{bbold}
\usepackage[toc]{appendix}
\usepackage{relsize}
\usepackage{leftidx}
\usepackage{bbm}
\usepackage{stackengine}
\usepackage{float}

\definecolor{lapislazuli}{rgb}{0.15, 0.38, 0.61}
\definecolor{YKblue}{rgb}{0.0, 0.18, 0.65}
\definecolor{carmine}{rgb}{0.81, 0.09, 0.13}
\definecolor{lavender}{rgb}{0.84, 0.79, 0.87}

\hypersetup{
	colorlinks=true,
	linkcolor=YKblue,
	linktoc=page,
	citecolor=blue,
	urlcolor=YKblue
}

\begin{document}
	
\title{The role of the exchange-Coulomb potential in two-dimensional electron transport}
	
    \author{J. L. Figueiredo}
    \email{jose.luis.figueiredo@tecnico.ulisboa.pt} 
    \author{J. T. Mendon\c{c}a}
    \affiliation{GoLP - Instituto de Plasmas e Fus\~{a}o Nuclear, Instituto
    Superior T\'{e}cnico, Universidade de Lisboa, 1049-001 Lisboa, Portugal}
            
    \author{H. Ter\c{c}as}
     \affiliation{GoLP - Instituto de Plasmas e Fus\~{a}o Nuclear, Instituto
    Superior T\'{e}cnico, Universidade de Lisboa, 1049-001 Lisboa, Portugal}
    \affiliation{Instituto Superior de Engenharia de Lisboa, Instituto Polit\'{e}cnico de Lisboa, Rua Conselheiro Em\'{i}dio Navarro 1, 1959-007 Lisboa, Portugal}

\begin{abstract}
We develop a quantum kinetic theory of two-dimensional electron gases in which exchange is treated self-consistently at the Hartree-Fock level and enters as a nonlocal, momentum-dependent field in phase space. By starting from the Coulomb Hamiltonian, we derive a Hartree--Fock--Wigner equation for the electronic Wigner function and obtain a closed fluid model with exchange-corrected pressure, force, and current. For a single layer, we show that exchange renormalizes the Fermi velocity and can drive a long-wavelength plasmonic instability at low densities. In coupled layers, the same framework predicts acoustic-optical mode coupling, and an instability forming long-lived charge-imbalance patterns that are not predicted by classical Vlasov and Boltzmann models. Finally, we apply the kinetic model to the Coulomb drag problem and show how exchange substantially enhances the drag resistivity in dilute GaAs double wells, quantitatively matching experimental observations. 
\end{abstract}

\maketitle

\section{Introduction}\label{sec:intro}

The dynamics of electron transport in low-dimensional conductors is typically formulated in terms of kinetic equations for a single-particle distribution in phase space. For clean two-dimensional electron gases (2DEGs) in semiconductor heterostructures or atomically thin materials, semiclassical Boltzmann and Vlasov equations provide the standard framework for describing collective charge dynamics, screening, and interlayer interactions~\cite{giuliani2008quantum,ando1982electronic}. In these formulations, electrons propagate with a band dispersion modified by static mean fields, while electron-electron interactions enter either through self-consistent electrostatic potentials or through collision integrals that encode quasiparticle scattering~\cite{rammer2011quantum,hwang2005transport}. In the weak-perturbation limit, these kinetic descriptions reduce to the random-phase approximation (RPA), which yields the linear density response and the collective excitation spectrum of low-dimensional Fermi gases~\cite{stern1967polarizability,chaplik1972possible}.

At low temperatures and moderate-to-low carrier densities, 2DEGs become strongly influenced by exchange and correlation effects~\cite{grinberg1987exchange,pudalov2015thermodynamic,shi2002droplet}. In this regime, the interaction parameter \(r_{s}\) is of order unity or larger and the exchange energy per particle represents a sizeable fraction of the Fermi energy~\cite{attaccalite2002correlation}. A number of equilibrium manifestations of this regime are now well established, including negative electronic compressibility~\cite{eisenstein1992negative}, anomalous quantum capacitance~\cite{xia2009measurement}, and Friedel-like oscillations in the screened potential of impurities~\cite{efros2012electron}. In double-layer systems, Coulomb drag measurements at low temperatures reveal a strong sensitivity of the drag resistivity to the detailed structure of the electron-electron interaction and to many-body renormalisations of the quasiparticle spectrum~\cite{gramila1991mutual,jauho1993coulomb,zheng1993coulomb}. These observations indicate that a kinetic description of dilute, low-temperature 2DEGs that neglects exchange is, in general, incomplete.

Most theoretical treatments of exchange in low-dimensional electron systems rely on either static or weakly dynamical approximations. For equilibrium problems, exchange is typically incorporated through density-functional or Thomas--Fermi--Dirac type schemes, which renormalise the equation of state and the static screening length~\cite{kohn1965self,dirac1930note,perdew1981self}. In dynamical settings, including plasmon propagation and Coulomb-drag configurations, exchange and correlations are most often accounted for through local-field factors that renormalise the RPA susceptibility and the effective interaction~\cite{calmels1995exchange,richardson1994dynamical}, or via spin-resolved extensions developed in the context of spin Coulomb drag~\cite{d2003spin,weber2005observation}. These frameworks are inherently restricted to near-equilibrium and weakly perturbed regimes, as exchange enters only through static renormalisations of response functions or collective-mode dispersions. As a result, the exchange field does not appear as a genuinely nonlocal, momentum-dependent force in the kinetic equation, nor does it evolve self-consistently with the full nonequilibrium distribution function~\cite{brodin2019hydrodynamic,zamanian2015contribution}. This limitation prevents such approaches from capturing regimes in which exchange directly alters local acceleration, phase-space transport, and nonlinear dynamics in strongly inhomogeneous or far-from-equilibrium electron gases~\cite{haas2022exchange,manfredi2021fluid,yadav2025solitons,kitayev2024non}.

In this work we develop a quantum kinetic description of two-dimensional electron gases that includes exchange at the mean-field level. Starting from the second-quantised Coulomb Hamiltonian confined to a 2D metallic sheet, we perform a Hartree--Fock decomposition and derive a closed evolution equation for the electronic Wigner function. The resulting kinetic equation has the structure of a Vlasov equation in which both the phase-space velocity and the force are renormalised by a nonlocal Fock potential, in addition to the usual Hartree term~\cite{bonitz2016quantum,melrose2020quantum}. The exchange potential is obtained self--consistently from the distribution function and depends explicitly on momentum, so that it has no classical analogue. As long as higher-order correlations beyond Hartree--Fock remain small, this formulation is well-suited to treat dynamical problems and provides a unified framework that connects quantum and semiclassical regimes in low-dimensional electron transport. Unlike local-field-factor approaches, our method retains the full momentum dependence and self-consistent evolution of the exchange field. This, in turn, permits a description of nonlinear and far-from-equilibrium regimes that lie beyond the domain of validity of linear-response theories.

By taking exchange corrections into account, we start by deriving a modified fluid equations for degenerate 2DEGs and analysing how exchange alters the propagation and stability of plasmons in isolated and coupled layers. For a single sheet, we obtain a degenerate plasmon dispersion where exchange appears as a negative correction to the Fermi velocity, enabling an exchange-driven instability in the low-density regime. Then, considering two spatially separated sheets, we derive the coupled dielectric response and show that exchange qualitatively changes the structure of acoustic and optical plasmon branches, opening a regime where both branches can become unstable and hybridise through acoustic-optical coupling. 

These analytical results are then complemented by numerical simulations of the full Hartree--Fock--Wigner kinetic equation which reveal several dynamical effects that cannot be captured by Hartree or RPA descriptions. We consider density modulations and identify clear parameter regimes in which exchange becomes comparable to, or even exceeds, the electrostatic contribution, thereby establishing the boundaries between electrostatic-dominated and exchange-dominated dynamics. We then examine dynamical screening and show that exchange fundamentally modifies the response of a dilute 2DEG, producing overscreening and oscillatory behaviour absent from the classical mean-field model. We further investigate coupled layers and demonstrate that exchange can destabilise the antisymmetric density mode, enabling the spontaneous emergence of long-lived spatial patterns of charge imbalance. Finally, we address a Coulomb drag configuration and find that exchange substantially enhances the drag resistivity at low temperatures, bringing the results into close agreement with experimental measurements in dilute GaAs bilayers. 

The remainder of the paper is organised as follows. In Sec.~\ref{sec_two} we derive the Hartree--Fock Hamiltonian for a 2D electron gas and obtain the corresponding quantum kinetic equation for the Wigner function. In Sec.~\ref{sec_three} we construct a fluid model with exchange corrections and obtain the dispersion relation of degenerate plasmons in a single sheet. Section~\ref{sec_four} extends the analysis to two spatially separated 2DEGs, deriving the coupled dielectric response and identifying exchange-modified acoustic and optical modes. In Sec.~\ref{sec_five} we present numerical solutions to the approximate Hartree--Fock--Vlasov kinetic equation, exploring local exchange effects, screening dynamics, and exchange-driven instabilities in double-layer geometries. Finally, Sec.~\ref{sec_conclusions} summarizes our main conclusions.

\section{Quantum model for the 2D electron gas}\label{sec_two}
Let us consider the conduction band of a two-dimensional metal sheet characterized by a large Fermi level $\mathcal E_F \gg k_BT$, with $T$ denoting the temperature. In this case, the conduction electrons behave as a one-component plasma, with the lattice providing a positive and neutralizing background. Hence the second-quantized Hamiltonian describing the electron system can be written as 
\begin{equation}
    \hat H = \sum_{n} \mathcal E_n \hat c_{ n}^\dagger   \hat c_{n} \  +\frac{1}{2} \sum_{n,m,n',m'} V_{n,m;n',m'} \hat c_{ n'}^\dagger \hat c_{ m'}^\dagger \hat c_{ m} \hat c_{ n} , \label{Hamiltonian}
\end{equation}
where
\begin{align}   
  & V_{n,m;n',m'} =  \nonumber \\
  &\int  d\mathbf r \int d\mathbf r' \ \phi^\ast_{n'}(\mathbf r)\phi^\ast_{m'}(\mathbf r') V(\mathbf r-\mathbf r') \phi_{m}(\mathbf r') \phi_{n}(\mathbf r)  
\end{align}
and $V(\mathbf r) = e^2 /(4\pi\varepsilon r)$ is the Coulomb potential. The single-particle states labeled by $n$ are solutions to the noninteracting problem associated with the metal sheet:
\begin{equation}
    \Big[ -\frac{\hbar^2}{2m}\bm{\nabla}^2 + U(z)\Big] \phi_n(\mathbf r) = \mathcal E_n \phi_n(\mathbf r) , \label{SPstates}
\end{equation}
with $U(z)$ the confining potential in the longitudinal direction. Since we assume free motion along $\mathbf r_\perp = (x,y)$, then each single-particle state is labelled by $n=(\mathbf k_{\perp}, \ell)$ where $\mathbf k_{\perp} = (k_x,k_y)$ is the two-dimensional wavevector associated with in-plane momentum eigenvalues and $\ell = 1,2,...$ denotes the subband due to longitudinal confinement. The corresponding states take the form 
\begin{equation}
    \phi_n(\mathbf r) = \frac{e^{i\mathbf k_{\perp}\cdot r_\perp}}{\sqrt{\mathcal A}} \xi_\ell(z),
\end{equation}
where $\xi_\ell(z)$ is the $\ell$-th confinement wavefunction and $\mathcal A$ the surface area of the sample. Consequently, $\mathcal E_n = \hbar^2 \mathbf k_\perp^2/(2m) +  \mathcal E_\ell$, with $m$ the effective mass determined by the lattice geometry and $\mathcal E_\ell \sim \ell^2/a^2$ for a quantum-well potential with thickness $a$. 

If we assume that $a$ is sufficiently small such that the energy spacing $\mathcal E_{\ell+1} - \mathcal E_{\ell}$ is large compared to the Fermi energy, then the electronic motion will be mainly restricted to the lowest subband. As a result, only $\ell=1$ intra-(sub)band scattering events are kept in Eq.~\eqref{Hamiltonian}. Moreover, taking the limit $a \rightarrow 0$ leads to $|\xi_{1}(z)|^2 = a\delta(z)$, so the Coulomb interaction becomes (we drop the $\ell=1$ index from now on)
\begin{equation}
    \frac{1}{2} \sum_{\mathbf k_\perp,\mathbf k'_\perp,\mathbf q_\perp}  V_{\mathbf q_\perp}^\text{(2D)} \hat c_{\mathbf k_\perp + \mathbf q_\perp}^\dagger \hat c_{\mathbf k'_\perp - \mathbf q_\perp}^\dagger \hat c_{\mathbf k'_\perp} \hat c_{\mathbf k_\perp} , \label{Coulomb_final}
\end{equation}
with $V_{\mathbf q_\perp}^\text{(2D)} = e^2/(2 \varepsilon \mathcal A q_\perp )$ being the two-dimensional Fourier transform of $V(\mathbf r_\perp)$. The lowest-subband energy can be set to zero without loss of generality. Moreover, we consider a nonmagnetized, unpolarized 2DEG governed by the spin-independent Coulomb interaction of Eq.~\eqref{Coulomb_final} and neglect any relativistic correction such as Zeeman splitting or spin-orbit coupling. In this situation the dynamics is identical for each spin component, so for notational simplicity we have suppressed the spin index in the intermediate steps. \\

 Next, we apply the Hartree-Fock decomposition to the Coulomb interaction in Eq.~\eqref{Coulomb_final}, as described in Ref.~\cite{PhysRevA.110.063519}. In practice, this procedure amounts to neglect higher-order quantum fluctuations of fermionic operators by keeping only the mean-field contributions plus a small correction associated with electron--electron collisions. We are led to the effective Hamiltonian  
\begin{equation}
      \hat H = \sum_{\mathbf k,\mathbf k'} \mathcal H_{\mathbf k,\mathbf k'} \hat c^\dagger_{\mathbf k'} \hat c_{\mathbf k} +  \hat{\mathcal C} , \label{HF_Hamiltonian}
\end{equation}
where 
\begin{equation}
 \mathcal   H_{\mathbf k,\mathbf k'} = \delta_{\mathbf k,\mathbf k'} \frac{\hbar^2 \mathbf k^2}{2m} +  \Phi^\text{H}_{\mathbf k' -\mathbf k} + \Phi^\text{F}_{\mathbf k,\mathbf k'} \label{Hk_k}
\end{equation}
contains the kinetic energy plus mean-field Coulomb interaction. The $\hat{\mathcal C}$ term translates collisional effects above mean-field, which provide $\mathcal{O}(e^4)$ corrections. The mean-field terms are separated into the Hartree (or \textit{electrostatic}) potential,
\begin{equation}
    \Phi^\text{H}_{\mathbf k} =  V_{\mathbf k}^\text{(2D)} \langle \hat n_{\mathbf k} \rangle, \label{HartreePot_k}
\end{equation}
with $\hat n_{\mathbf k} = \mathcal A \sum_{\mathbf q} \hat c^\dagger_{\mathbf q} \hat c_{\mathbf k + \mathbf q}$ denoting the density operator in Fourier space, and the Fock (or \textit{exchange}) potential,
\begin{equation}
    \Phi^\text{F}_{\mathbf k,\mathbf k'} = - \sum_{\mathbf q}  V_{\mathbf q}^\text{(2D)} \langle \hat c^\dagger_{\mathbf k + \mathbf q} \hat c_{\mathbf k' + \mathbf q} \rangle.
\end{equation}
The latter arises due to the fermionic anticommutation relations and has no classical analogue. 

In what follows we restrict attention to regimes where electron-electron correlations beyond mean field are subleading and neglect the collision operator $\hat{\mathcal C}$. This approximation is justified for dilute, low-temperature 2DEGs in which the characteristic plasmon and drift frequencies exceed the electron-electron scattering rate, so that the dynamics is dominated by long-wavelength, nearly collisionless collective motion~\cite{bonitz2016quantum}. Under these conditions, Eqs.~\eqref{HF_Hamiltonian} and~\eqref{Hk_k} define a closed evolution equation for the electronic Wigner function $f_{\mathbf k}(\mathbf r) \equiv \sum_{\mathbf k'} e^{i \mathbf k'\cdot \mathbf r} \langle \hat c_{\mathbf k - \mathbf k'/2}^\dagger \hat c_{\mathbf k + \mathbf k'/2} \rangle$~\cite{PhysRevA.110.063519}, which reads 
\begin{align}
    &\frac{\partial}{\partial t} f_{\mathbf k}(\mathbf r)  =  \nonumber  \\
    &\frac{2}{\hbar} \mathcal H_{\mathbf k}(\mathbf r)\sin\left(\frac{1}{2} \frac{\overleftarrow{\partial}}{\partial \mathbf{r}} \cdot \frac{\overrightarrow{\partial}}{\partial \mathbf{k}} - \frac{1}{2} \frac{\overleftarrow{\partial}}{\partial \mathbf{k}} \cdot\frac{\overrightarrow{\partial}}{\partial \mathbf{r}}   \right) f_{\mathbf k}(\mathbf r). \label{KinEq}
\end{align}
In particular, we use the spin-summed Wigner distribution $f_{\mathbf{k}}(\mathbf{r},t)=\sum_{\sigma=\uparrow,\downarrow} f_{\mathbf{k}\sigma}(\mathbf{r},t)$ so that phase-space moments of $f_{\mathbf{k}}$ directly yield the physical density and current of an unpolarized 2DEG. Extensions to the polarized/magnetized case would require evolving the spin-resolved distributions $f_{\mathbf{k}\uparrow}$ and $f_{\mathbf{k}\downarrow}$ and adding the corresponding spin-dependent terms to the Hamiltonian.

In Eq.~\eqref{KinEq}, $\mathcal H_{\mathbf k}(\mathbf r) = \sum_{\mathbf k'} e^{i\mathbf r \cdot \mathbf   k'}\mathcal  H_{\mathbf k - \mathbf k'/2,\mathbf k + \mathbf k'/2}$ is the Wigner transform of $\mathcal H_{\mathbf k,\mathbf k'}$, given by
\begin{equation}
    \mathcal H_{\mathbf k}(\mathbf r) = \frac{\hbar^2 \mathbf k^2}{2m} + \Phi^\text{H}(\mathbf r) + \Phi^\text{F}(\mathbf r,\mathbf k).
\end{equation}
The Hartree potential $\Phi^\text{H}(\mathbf r) = \mathcal A^{-1} \sum_{\mathbf k} e^{i\mathbf r \cdot \mathbf k }  \Phi^\text{H}_{\mathbf k}$ is related with the distribution function through Poisson's equation~\footnote{Equation~\eqref{HartreePot_k} is equivalent to Poisson's equation. This can be shown by writing $\Phi^\text{H}(\mathbf r) = \frac{1}{\mathcal A}  \sum_{\mathbf k} e^{i\mathbf r \cdot \mathbf k }  \Phi^\text{H}_{\mathbf k} = \frac{1}{\mathcal A}\sum_{\mathbf k} e^{i\mathbf r \cdot \mathbf k }  \frac{\langle n_{\mathbf k}\rangle }{2\varepsilon \mathcal A |\mathbf k|}$ and using the identity $\frac{1}{|\mathbf r|} =\frac{1}{\mathcal A} \sum_{\mathbf k} e^{i\mathbf k\cdot \mathbf r } \frac{2\pi}{|\mathbf k|}$ and $\langle n_{\mathbf k}\rangle = \mathcal A \int d\mathbf r \ e^{-i\mathbf k\cdot \mathbf r} n(\mathbf r)$, with $n(\mathbf r)$ the real-space density. Upon performing the $\mathbf k$ summation we are led to $\Phi^\text{H}(\mathbf r) = \frac{e^2}{4\pi \varepsilon } \int d\mathbf r' \frac{n(\mathbf r')}{|\mathbf r-\mathbf r'|}$. The Poisson equation is now evidenced by making use of the relation $\bm{\nabla}^2 \frac{1}{|\mathbf r-\mathbf r'|} = -4\pi \delta(\mathbf r-\mathbf r')$, leading to $\bm{\nabla}^2 \Phi^{H}(\mathbf r) = - e^2 n(\mathbf r)/\varepsilon$}, whereas the Fock potential verifies the self-consistent condition
\begin{equation}
     \Phi^\text{F}(\mathbf r,\mathbf k) = -\sum_{\mathbf k'}  V_{\mathbf k'}^\text{(2D)} f_{\mathbf k + \mathbf k'}(\mathbf r). \label{ExPot}
\end{equation}
Contrarily to the Hartree potential, the value of $\Phi^\text{F}$ at a given spatial position $\mathbf r$ only depends on the value of the distribution function at $\mathbf r$. Hence we can think of the Fock potential as a dynamical term responsible for preventing any two electrons from occupying the same eigenstate, which explains its functional dependence on the distribution function, and in particular, its $\mathbf k$ dependence. Moreover, when $f_{\mathbf k}(\mathbf r)$ vanishes at a given region $S$ of real space, the local energy $\mathcal H_{\mathbf k}(\mathbf r)$ meausured at $S$ has only contribution from the kinetic plus Hartree terms. This means that a test particle occupying a real-space volume that contains no other particle will feel no exchange potential since the exclusion principle is automatically respected in that case. \par 

\section{Fluid model with exchange corrections}\label{sec_three}
The kinetic equation \eqref{KinEq} describes mean-field effects at all length scales and is valid as long as electron collisions can be neglected. The quantum-mechanical effects are included in the differential operator therein, containing an infinite number of spatial derivatives that account for quantum uncertainty in phase space. When the distribution function varies slowly in space, a semiclassical approximation can be applied by retaining only the first term in the differential series. The result is a modified Vlasov equation which includes renormalizations of phase-space velocity and force promoted by the exchange potential. \par 
After establishing the semiclassical kinetic equation, it is convenient to take its moments and derive a set of fluid equations for macroscopic quantities. The first two of these quantities are the density and kinetic current, $n(\mathbf r,t) = \mathcal A^{-1}\sum_{\mathbf k} f_{\mathbf k}(\mathbf r)$ and $\bm{j} (\mathbf r,t)= \mathcal A^{-1}\sum_{\mathbf k} (\hbar \mathbf k/m )f_{\mathbf k}(\mathbf r)$, respectively. From Eq.~\eqref{KinEq} we obtain 
\begin{align}
&\frac{\partial}{\partial t} n + \bm\nabla\cdot \Big( \bm{j} +  \bm{j}^\text{F}\Big)  = 0 , \label{dnDt} \\
&\frac{\partial}{\partial t}\bm{j} +\frac{1}{m}\bm\nabla \Big( P + P^\text{F} \Big) =  \frac{n}{m}\Big( -\bm\nabla \Phi^\text{H} + \bm{f}^\text{F} \Big).  \label{djdt}
\end{align}
The set of fluid variables $\bm{j}^\text{F}, P^\text{F}$ and $ \bm{f}^\text{F}$ arise due to the Fock potential and can be interpreted as degenerate contributions to the current, pressure and force, respectively, steaming from the velocity and force renormalizations provided by $\Phi^\text{F}$. In particular, $ \bm{f}^\text{F}$ is the exchange counterpart to the electrostatic potential, while $\bm{j}^\text{F}$ and $P^\text{F}$ have no electrostatic analogue. They are determined by
\begin{align}
    \bm{j}^\text{F}(\mathbf r,t) &= \frac{1}{\mathcal A} \sum_{\mathbf k}  \frac{1}{\hbar}\frac{\partial  \Phi^\text{F}}{\partial \mathbf k}f_{\mathbf k } (\mathbf r),  \\
    P^\text{F}(\mathbf r,t) &= \frac{1}{\mathcal A} \sum_{\mathbf k}  \Bigg( \frac{\partial \Phi^\text{F}}{\partial \mathbf k}\otimes \mathbf k \Bigg)f_{\mathbf k } (\mathbf r),\\
     \bm{f}^\text{F}(\mathbf r,t) &= -\frac{1}{n\mathcal A} \sum_{\mathbf k} \frac{\partial  \Phi^\text{F}}{\partial \mathbf r}f_{\mathbf k } (\mathbf r) .
\end{align}
Moreover $P$ is the usual kinetic pressure, 
\begin{equation}
	P (\mathbf r,t) = \frac{1}{\mathcal A} \sum_{\mathbf k} \frac{\hbar \mathbf k \otimes \hbar \mathbf k}{m} f_{\mathbf k}(\mathbf r).
\end{equation}
Using a similar procedure, equations of motion for the remaining macroscopic variables can also be established from Eq.~\eqref{KinEq}, leading to an infinite chain of coupled equations for higher-order fluid moments. A truncation procedure is thus necessary to close the system after relating the higher-order moments with density and current. 

\subsection{Equations of state}
Equations of state can be derived by assuming an ansatz for the distribution function that is locally valid for the entire evolution. For room temperature conditions of a 2DEG with large Fermi temperature, the appropriate function for our case of interest corresponds to the $T\rightarrow 0$ limit of a displaced Fermi-Dirac function, 
\begin{equation}
    f^{(0)}_{\mathbf k}(\mathbf r) = 2 \Theta\big(k_F - |\mathbf k - \mathbf k_0|\big), \label{HS}
\end{equation}
where $\mathbf k_0 \equiv \mathbf k_0 (\mathbf r,t)$ is the average wavevector, $k_F \equiv \sqrt{2\pi n(\mathbf r,t)}$ is the local Fermi wavevector and $\Theta(x)$ is the Heaviside-step function. Additionally, the factor of two accounts for the spin degeneracy. One can verify that $\mathcal A^{-1} \sum_{\mathbf k} f^{(0)}_{\mathbf k} = n$ and $\mathcal A^{-1} \sum_{\mathbf k} \mathbf k f^{(0)}_{\mathbf k} = n\mathbf k_0$ as expected. \par
\begin{figure}
\centering \hspace{-0.7cm}
\includegraphics[scale=0.55]{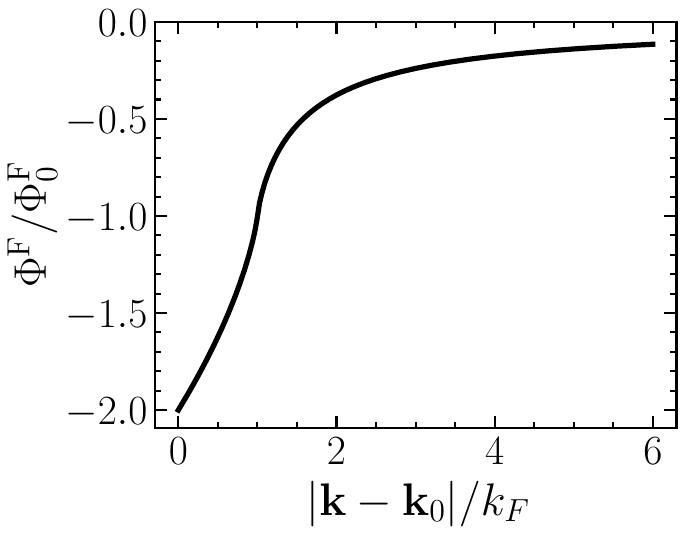}
\caption{Normalized Fock potential of Eq.~\eqref{ExchH}.}
\label{FockP}
\end{figure}

 
After replacing Eq.~\eqref{HS} into Eq.~\eqref{ExPot} and performing the integrations, we are led to
\begin{equation}
    \Phi^\text{F}(\mathbf r,\mathbf k) = - \Phi_0^\text{F}  \ G\Big(\frac{k_F}{|\mathbf k - \mathbf k_0|}\Big) \label{ExchH},
\end{equation}
with $ \Phi_0^\text{F} = e^2 k_F/(\pi^2 \varepsilon)$ corresponding to the overall scale and $G(x) = 1 + \log\big[(1+\sqrt{x})/(\abs{1-\sqrt{x}})\big](x-1)/(2\sqrt{x})$ a function of order unit, shown in Fig.~\ref{FockP}. The (negative) sign of $\Phi^\text{F}$, combined with the fact that $\Phi_0^\text{F}$ is a growing function of the density, is behind the tendency of exchange-dominated systems to locally increase the density as equilibrium is reached. This is in accordance with previous works that have considered the exchange contribution to the energy in three-dimensional plasmas \citep{PhysRev.111.442,PhysRev.121.941,PhysRevA.110.063519}. An order-of-magnitude estimate of the importance of exchange is given by the ratio $\gamma = \Phi_0^\text{F}/\mathcal E_F$, where $\mathcal E_F = \hbar^2 k_F^2/2m$ is the Fermi energy. Since $\gamma \sim n^{-1/2}$, we expect that for high carrier densities the kinetic energy dominates, which defines the ballistic regime. When the density decreases below a certain threshold value, exchange effects surpass kinetic energy and the dynamics becomes dominated by degeneracy. \par 
From previous results, the following equations of state can be derived
\begin{align}
    P &= \frac{   \hbar^2 \pi n^2}{2m} \delta_{i,j}  + \frac{m}{n} \bm{j}\otimes \bm{j},\\
    P^\text{F} &= \frac{3\sqrt{2}e^2 \sqrt{n}}{20 \pi^{3/2} \varepsilon}\delta_{i,j}, \\
    \bm{f}^\text{F} &= \frac{\sqrt{2} e^2}{\pi^{3/2} \varepsilon} \frac{1}{\sqrt{n}}\bm{\nabla} n , \label{fF}
\end{align}
together with $\bm{j}^\text{F} = 0$. Inserting these expressions into Eqs.~\eqref{dnDt} and \eqref{djdt} establishes a closed system of fluid equations. 

\subsection{Degenerate plasmon dispersion}
The dispersion relation of collective plasma oscillations (or \textit{plasmons}) can be found after linearizing the fluid model and transforming each equation to Fourier space. The result is a the dielectric function $\epsilon(\omega,\mathbf k) = 1  + \chi(\omega,\mathbf k)$ with $\chi(\omega,\mathbf k)$ the susceptibility,
\begin{equation}
     \chi(\omega,\mathbf k) = -\frac{\mathcal U(\mathbf k)}{(\omega-\mathbf u_0 \cdot \mathbf k)^2 - U^2 k^2} \label{Pol}.
\end{equation}
Above, $\mathcal U(\mathbf k) = e^2 n_0 k/(2\varepsilon m)$, $\mathbf{u}_0 = \hbar \mathbf{k}_0/m$ is the drift velocity and $U^2 =  v_F^2/2 - \widetilde{v}^{\,2}$ contains the effect of kinetic energy through $v_F = \hbar k_F/m$ and exchange energy through $\widetilde{v}$. The latter reads
\begin{equation}
   \widetilde{v}  = \sqrt{\frac{17 e^2 k_F }{20\pi^2  \varepsilon m  }} , \label{vtilde}
\end{equation}
and renormalizes the Fermi velocity with a negative contribution. A similar effect has been found in three dimensional plasmas using different methods \citep{PhysRevE.92.013104}.\par 
The roots of $\epsilon(\omega,\mathbf k)$ determine the dispersion relation of collective modes, providing
\begin{equation}
    \omega(\mathbf k) = \mathbf u_0 \cdot \mathbf  k \pm \sqrt{\alpha k + U^2 k^2} ,\label{plasmonDisp}
\end{equation}
with $\alpha = e^2 k_F^2/(4\pi \varepsilon m)$. Taking $ \widetilde{v} = 0$ recovers the result of Refs.~\citep{PhysRevB.23.805,PhysRevB.39.6208} for the long-wavelength limit, where collective modes were determined using the Random Phase Approximation (RPA) to calculate an approximate susceptibility function $\chi^\text{RPA}$. Formally, $\chi^\text{RPA}$ contains the dispersion to any order in $k$ but neglects the Fock potential. Such result can also be found from Eq.~\eqref{KinEq} by setting $\Phi^\text{F}=0$. However, since we consider the exchange energy, our method is able to predict a lower-order correction (at order $k^2$) to the plasmon dispersion that is \textit{not} included in $\chi^\text{RPA}$.

An immediate consequence of the negative exchange contribution $-\widetilde{v}$ is that nonzero imaginary parts of the frequency are now possible even for a single-sheet configuration if $U^2<0$, which requires $n_0 \lesssim 10^{14} \, \text{cm}^{-2}$ for typical GaAs effective mass $m = 0.067 m_e$. Denoting $\omega(\mathbf k)= \Omega(\mathbf k) + i\gamma(\mathbf k)$ and assuming that $U^2$ is negative, we get a finite value of $\gamma$ for $k > k_c$, with $k_c = \alpha/|U^2|$ the lowest unstable mode. For small displacements $\delta k = k - k_c$, it follows that
\begin{equation}
    \gamma \simeq \pm \alpha^{1/2} \Bigg( \delta k^{1/2} + \frac{|U|}{2  \alpha}  \delta k^{3/2}\Bigg) . \label{imgP}
\end{equation}

\begin{figure*}
    \centering
    \includegraphics[scale=0.6]{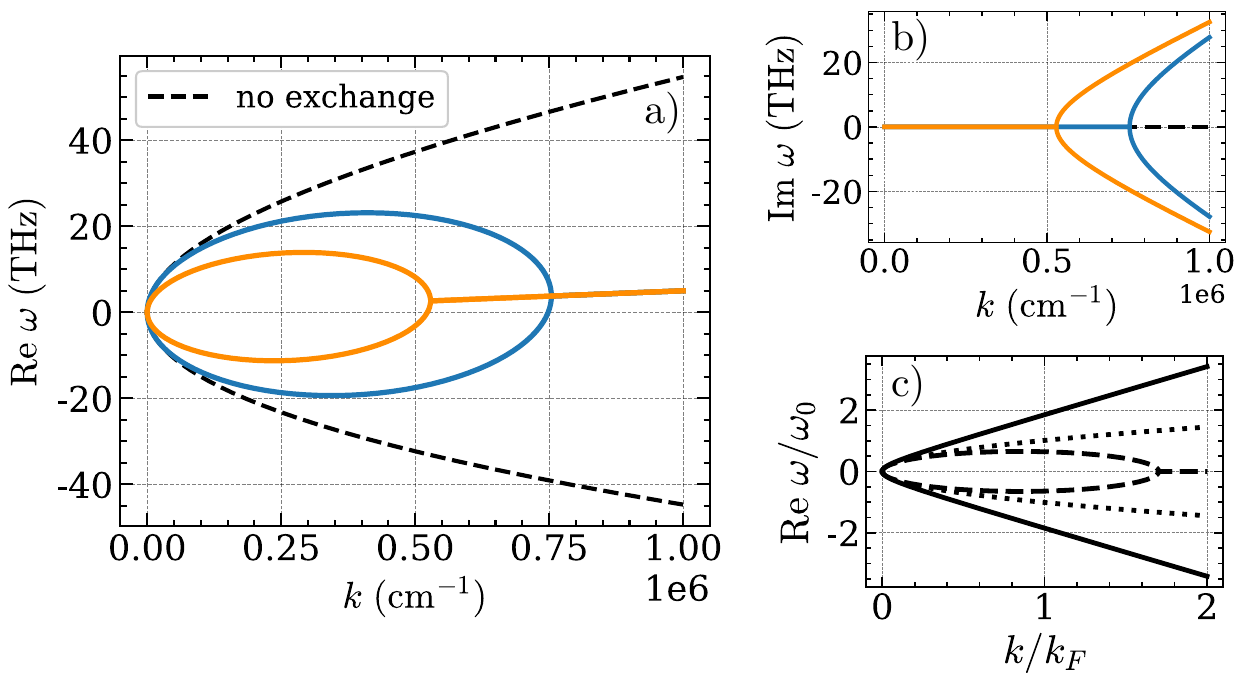}
    \caption{Real (a) and imaginary (b) parts of the plasmon dispersion including the effect of exchange (solid curves) and neglecting exchange (dashed curve) for varying densities $n_0 = 10^{10}\, \text{cm}^{-2}$ (orange) and $n_0 = 10^{11} \,\text{cm}^{-2}$ (blue) with equilibrium  velocity $u_0 = 5\times 10^{4} \,\text{m}\text{s}^{-1}$. (c): real part of the plasmon dispersion for growing values of the density. The solid curve corresponds to the critical density ($n_c \sim 2\times 10^{14}\, \text{cm}^{-2}$) above which the dispersion is real. (typical GaAs parameter $m = 0.067 m_e$) }
    \label{plasmon_disp}
\end{figure*}

These exchange instabilities result from a competition between kinetic and exchange pressures, and become more important for larger Coulomb energies. In Fig.~\ref{plasmon_disp} [panel (a)] the real part of the dispersion is plotted along with the result in the absence of exchange ($\widetilde{v}=0$). In panel (b) the corresponding imaginary parts are shown, which follow Eq.~\eqref{imgP} close to the origin. Panel (c) shows the real part of the dispersion in the case of zero drift in both the regime dominated by the kinetic effects (higher densities) as well as in the regime dominated by exchange (lower densities).

\section{Stream instabilities in parallel sheets}\label{sec_four}
After establishing the dispersion relation of degenerate plasmons for isolated metal sheets, let us now focus on the consequences for two-stream plasmonic instabilities. In particular, we are interested in the \textit{Coulomb drag} configuration, where two metallic plates initially in equilibrium with finite currents are brought sufficiently close to each other. \par 
Assuming that electron-electron correlations between different plates are unimportant, we may treat each electron gas as an independent fluid and thus associate to each plate a different distribution function $f^{(i)}_{\mathbf k}(\mathbf r)$, with $i=1,2$. We consider that plate $i$ is located at $z=Z_i$ and lies perpendicular to the $z$ direction such that $f^{(i)}_{\mathbf k}(\mathbf r)\sim \delta(z-Z_i)$, with $d = |Z_1 - Z_2|$ denoting the distance between the two plates. The distribution functions establish the density and current of each fluid, $n_{i}$ and $\bm j_{i}$, which also contain a factor of $\delta(z-Z_i)$. Since the Fock potential is spatially local, electrons in different fluids only interact through electrostatic forces, while electrons belonging to the same fluid interact both through electrostatic and exchange potentials. Denoting by $\Phi^\text{H}(\mathbf r_\parallel,Z_i) \equiv \Phi^\text{H}_i(\mathbf r_\parallel)$ the Hartree potential at each plate and using Poisson's equation, we get for the Fourier transform
\begin{align}
    \Phi^\text{H}_{i,\mathbf k_\parallel} &= \frac{e^2}{2\varepsilon k_\parallel} \Big[n_i(\mathbf k_\parallel)  + n_{\overline{i}}(\mathbf k_\parallel) e^{-d k_\parallel }\Big], \label{Poisson_CD}
\end{align}
where $\overline{i}$ corresponds to the plate contrary to $i$. The remaining fluid terms follow from the previous section upon establishing the equilibrium of each plate. We assume that the ansatz of Eq.~\eqref{HS} is valid for both plates (using $\mathbf k_{0,1}$ and $\mathbf k_{0,2}$, respectively).

After establishing the fluid model for each plate and taking into account the coupling promoted by the electrostatic potential given by Eqs.~\eqref{Pol} and~\eqref{vtilde}, we are led to the dielectric function for the coupled system (dropping again the parallel subscript)
\begin{align}
    \epsilon(\omega,\mathbf k) &= 1 + \chi_1(\omega,\mathbf k) + \chi_2(\omega,\mathbf k) \nonumber \\
    &+ \chi_1(\omega,\mathbf k)\chi_2(\omega,\mathbf k)\big(1-e^{-2kd}\big), \label{dispCD}
\end{align}
with each $\chi_i$ being that of Eq.~\eqref{Pol} with the appropriate $\mathbf u_0$ for each plate. \par 
As $d\rightarrow \infty$ the coupling between plates vanishes and $\epsilon(\omega,\mathbf k)$ becomes separable. Then, the collective modes are those of Eq.~\eqref{plasmonDisp} for independent metal plates. \par  
\subsection{Isotropic equilibrium}
In the case of zero macroscopic current and equal Fermi levels, the roots of Eq.~\eqref{dispCD} admit the following solution
\begin{equation}
    \omega_{\pm}^2 = U^2 k^2 + \alpha k\Big(1\pm e^{-dk}\Big). \label{ZeroUModes}
\end{equation}
For sufficiently small $d$, these reduce to optical $\omega_+ \simeq \pm \sqrt{\alpha k}$ and acoustic $\omega_- \simeq \pm \sqrt{U^2} k$ modes.  Clearly if $U^2>0$ no instability can occur for $\mathbf u_{0,i}=0$. Remarkably, in the presence of sufficiently strong exchange, the acoustic branch becomes unstable even at $\mathbf u_{0,i}=0$, with growth rate $\gamma = |U| k$, which is different from $\gamma = \sqrt{\alpha k }$ obtained in Eq.~\eqref{imgP} for isolated plates. For typical values, we can verify that $|U| k \gg \sqrt{\alpha k }$, which means that the electrostatic interaction between the two fluids leads to an increase of the growth rate even in the absence of streaming as long as $d>1/k_F \sim \text{nm}$ and $n<10^{14}\, \text{cm}^{-2}$ for typical GaAs parameters.

\subsection{Finite equilibrium currents}
Now let us consider the stream instability in the presence of stationary currents. These can be achieved by applying a voltage. In this case we expect modifications to the optical and acoustic modes of Eq.~\eqref{ZeroUModes} for sufficiently small $d$. Our goal is to understand the impact of exchange effects in the growth rates of collective modes. Noting that Eq.~\eqref{Pol} determines that modes propagating perpendicularly to the either $\mathbf u_{0,1}$ or $\mathbf u_{0,2}$ admit no instability if the effect of exchange is neglected, we focus on the case of parallel $\mathbf u_{0,1}$ and $\mathbf u_{0,2}$ and consider modes propagating along the same direction.\par 
Previous works considering these stream instabilities show that, when $d$ is sufficiently small, the acoustic branch becomes unstable. On the other hand, the optical branch was found to be always stable, independently of the value of $d$ ~\cite{PhysRevB.23.805,PhysRevB.39.6208}. After including exchange effects, we find that also the optical plasmon branch acquires an imaginary part due to acoustic-optical coupling. Additionally, the growth rate of acoustic modes is also modified. \par
\begin{figure}
    \centering
    \includegraphics[scale=0.65]{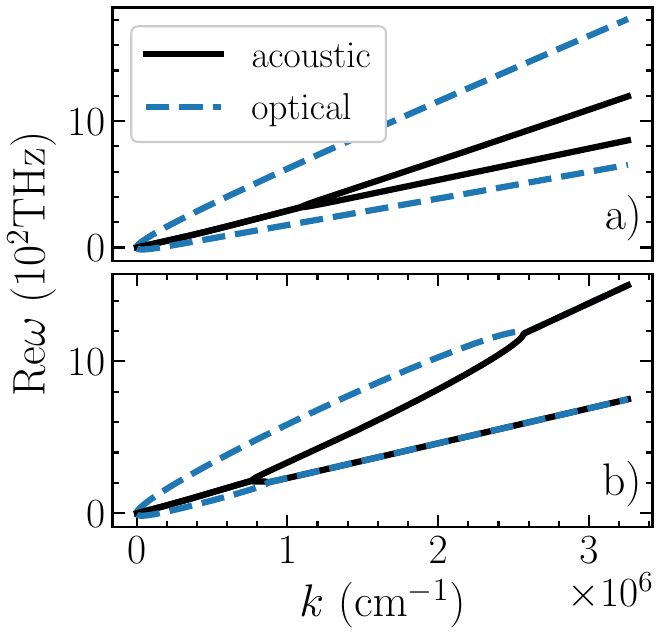}
    \caption{Acoustic and optical plasmons (a) if exchange is neglected and (b) if exchange is considered, the latter displaying acoustic-optical mode coupling ($n_{0,2}/n_{0,1} = 0.1$, $u_{0,2}/u_{0,1} = 0.5$, $d=10\,$nm.)}
    \label{plasmons_CD}
\end{figure}

\begin{figure}
    \centering
    \includegraphics[scale=0.7]{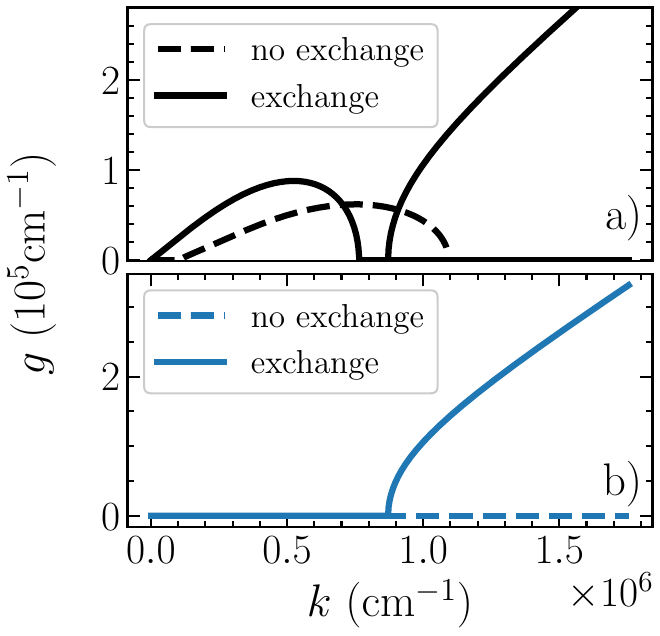}
    \caption{Comparison between the gains of (a) lower acoustic and (b) lower optical modes excluding and including exchange effects. The simulation parameters are those of Fig.~\ref{plasmons_CD}.}
    \label{gain_modes}
\end{figure}

 
Figure~\ref{plasmons_CD} shows the real parts of all four modes when $d=1\,\textup{\AA}$ and compares the cases where exchange is both considered and neglected. In the absence of exchange, acoustic modes hybridize, leading to a growth rate that has been described in previous works. When exchange is included, we observe that, apart from the acoustic coupling, additional acoustic-optical hybridizations arise. As a consequence, the gain $g(k) = \gamma(k)/v_k$ (with $v_k$ denoting the phase velocity) is modified for both acoustic and optical modes. This quantity is plotted in Fig.~\ref{gain_modes} using the same parameters as Fig.~\ref{plasmons_CD} for comparison. In particular, panel \ref{gain_modes}{\color{YKblue}.b)} shows that a finite gain is attained for the optical branch, which can be fully attributed to exchange interactions. \par

\section{Numerical simulations of degenerate dynamics}\label{sec_five}
To get further insights on the nonlinear dynamical effects promoted by exchange, the fluid model derived before is not enough and the full kinetic equation \eqref{KinEq} must be considered. For sufficiently slow spatial variations, it takes the form of a Vlasov equation with exchange corrections,
\begin{align}
    &\Bigg[ \frac{\partial}{\partial t} + \bm{ V}_{\mathbf k}(\mathbf r) \cdot \frac{\partial}{\partial \mathbf r} + \frac{1}{\hbar}\bm{ F}_{\mathbf k}(\mathbf r) \cdot \frac{\partial}{\partial \mathbf k} \Bigg]f_{\mathbf k}(\mathbf r)  = 0 ,\label{KinEq2}
\end{align}
with
\begin{align}
   \bm{ V}_{\mathbf k}(\mathbf r) &= \frac{\hbar \mathbf k}{m} + \frac{1}{\hbar}  \frac{\partial}{\partial \mathbf k}    \Phi^\text{F}(\mathbf r,\mathbf k), \label{VFv}\\
    \bm{F}_{\mathbf k}(\mathbf r) &= - \frac{\partial}{\partial \mathbf r}    \Phi^\text{H}(\mathbf r) - \frac{\partial}{\partial \mathbf r}    \Phi^\text{F}(\mathbf r,\mathbf k), \label{VFf} 
\end{align}
denoting, respectively, the phase-space velocity and force. Besides the first (classical) term, Eqs.~\eqref{VFf} and~\eqref{VFv} also include a contribution from exchange. Because $\Phi^\text{F}$ also depends on the distribution function through Eq.~\eqref{ExPot}, these corrections introduce nonlinear contributions which may be important not only to the onset of plasmonic instabilities but also to the saturation regime and long-time dynamics. 

The Hartree--Fock--Vlasov kinetic equation~\eqref{KinEq2} is accurate when the dynamics is governed by long-wavelength, nearly collisionless excitations, for which correlation effects beyond mean field give only subleading corrections. This applies to the dilute, low-temperature 2DEGs considered here, where plasmon and drift frequencies exceed electron--electron scattering rates and spatial variations occur on scales of order $k_{F}^{-1}$ or larger. In this regime, exchange enters as the leading correction to Vlasov or RPA-based descriptions and controls the nonequilibrium phenomena studied below.

For the sake of numerical tractability, we restrict the spatial dynamics to one dimension, $f_{\mathbf{k}}(\mathbf{r}) \rightarrow f_{k}(x)$, while the Hartree and Fock potentials are still evaluated using the full 2D Fourier kernels, ensuring that the correct screening and exchange physics of a two-dimensional electron gas is preserved. Additionally, all quantities are recast in dimensionless form through the transformations \(x \rightarrow k_{F} x\), \(k \rightarrow  k/k_{F}\), \(V \rightarrow   V / v_F\), \(t \rightarrow  \omega_F t \), and \(F \rightarrow  F / (k_F \mathcal E_F)\), $v_F = \hbar k_F/m$ being the Fermi velocity, $\omega_F = \mathcal E_F/\hbar$ the Fermi frequency, and $
\mathcal E_F = \hbar^2 k_F^2/(2m)$ the Fermi energy. With this choice, the Coulomb terms depend only on the dimensionless Wigner-Seitz radius $r_s = e^2 m/(\hbar^2 k_F \varepsilon)$, which controls the relative strength of interactions. Unlike the three-dimensional electron gas, where the large-density limit \(k_{F} \rightarrow \infty\) does not suppress interactions, in two dimensions the Coulomb contribution vanishes when \(n \rightarrow \infty \) defining the ballistic regime. In the following we focus on intermediate values \(r_s \sim 1\), corresponding to typical carrier densities \(n \sim 10^{10}\text{--}10^{12}\,\mathrm{cm}^{-2}\) of GaAs quantum wells. 

The remainder of this section explores the physical consequences of the exchange terms in Eq.~\eqref{KinEq2} for the dynamics of degenerate electron systems. Our results are based on numerical solutions of the kinetic equation obtained with a semi-Lagrangian scheme for phase-space advection~\cite{crouseilles2010conservative}, finite differences for spatial and momentum gradients, and fast Fourier transforms to compute the Hartree and Fock fields. Initial conditions are propagated with a Strang-splitting integrator~\cite{einkemmer2018low}, which provides stable access to the strongly nonlinear regime while preserving the fine-scale structure of the distribution function.

\subsection{Local exchange effects}\label{sec_local_exchange}

Before addressing the full Hartree-Fock electronic dynamics, it is convenient to isolate the role of exchange in a simple setting in order to understand its main features. In particular, we would like to answer the practical question of when can the Fock terms be neglected in comparison with their electrostatic counterpart. To this end we consider a single 2DEG without streaming, described by the kinetic equation~\eqref{KinEq2}.

Let us assume that at $t=0$ the electron fluid is in a spatially modulated equilibrium of the form
\begin{equation}
f_k(x) = f_{0}(k;n_{0},T)\,\bigl[1 + \Delta\cos(2\pi x/\lambda)\bigr],
\end{equation}
where $f_{0}$ is a Fermi--Dirac distribution of density $n_{0}$ and temperature $T$, and $\Delta \ll 1$ is the modulation amplitude. For each choice of the modulation wavelength $\lambda$ and temperature we compute the self--consistent Hartree and Fock potentials and evaluate the corresponding forces, 
\begin{align}
    F^\text{H}(x)&= -\frac{\partial}{\partial x}\Phi^{H}(x), \\
    F^\text{F}(x)&= - \frac{1}{n(x)}\int \frac{d\mathbf k}{(2\pi)^2} \, \frac{\partial}{\partial x}\Phi^{F}(x,k) \,f(x,k),\label{Feff}
\end{align}
where $n(x) = \int  \frac{d\mathbf k}{(2\pi)^2} \,f(x,k)$ is the local density. Equation~\eqref{Feff} reduces to the exchange--fluid force of Eq.~\eqref{fF} in the long--wavelength limit. As a compact measure of the relative importance of exchange, we define the spatially averaged ratio
\begin{equation}
R(\lambda,T) =
\Bigl\langle
\frac{\bigl|F^\text{F}(x)\bigr|}
     {\bigl|F^\text{H}(x)\bigr|}
\Bigr\rangle_{x}.
\end{equation}

\begin{figure}
    \centering
    \includegraphics[scale=0.62]{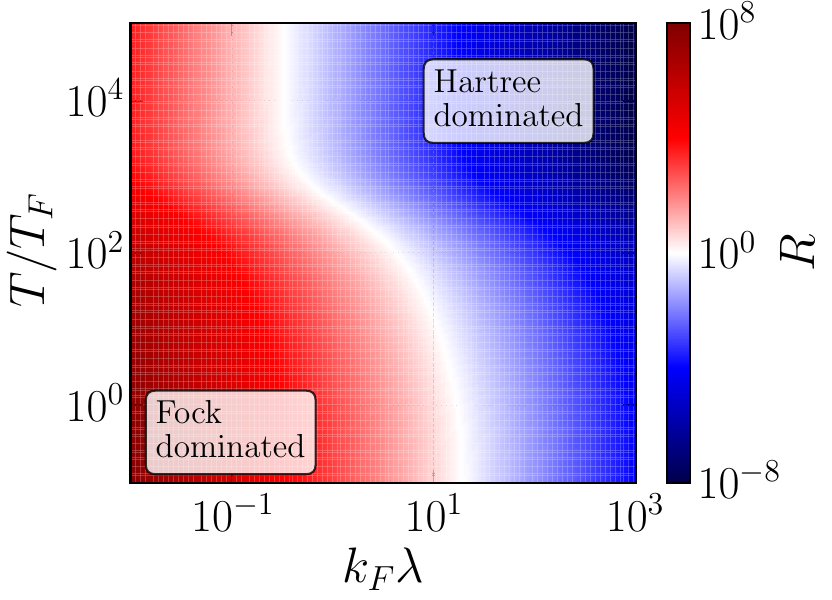}
    \caption{Mean ratio between the effective Fock and Hartree forces,
$R = \langle |F^{F}| / |F^{H}| \rangle_{x}$, as a function of the dimensionless modulation wavelength $k_{F}\lambda$ and reduced temperature $T/T_{F}$. Exchange is negligible in the upper left blue region ($k_{F}\lambda \gg 1$ or $T \gg T_{F}$), and becomes comparable to or larger than the Hartree force in the red region where $k_{F}\lambda \lesssim 1$ and $T \lesssim T_{F}$. }
\label{fig_HF_ratio}
\end{figure}

Figure~\ref{fig_HF_ratio} shows $R(\lambda,T)$ as a function of the dimensionless wavelength $k_{F}\lambda$ and reduced temperature $T/T_{F}$. The color map reveals two clear conditions for exchange to be relevant. First, for slowly varying density profiles with $k_{F}\lambda \gg 1$ the ratio $R$ is much smaller than unity for all temperatures, such that the dynamics is essentially Hartree dominated and a classical fluid description is adequate. Second, as the modulation wavelength approaches the Fermi wavelength, $k_{F}\lambda \lesssim 1$, the ratio rapidly increases and can exceed unity, signalling that exchange becomes comparable to or larger than the electrostatic force. This exchange--dominated regime is further confined to degenerate conditions $T \lesssim T_{F}$ because at higher temperatures, thermal smearing of the Fermi surface suppresses the Fock contribution even for short--wavelength perturbations. The map in Fig.~\ref{fig_HF_ratio} provides a simple phase diagram for the strength of exchange, showing that only short--scale, low--temperature density modulations probe the full quantum character of exchange forces.

\begin{figure}
    \centering
    \includegraphics[scale=0.6]{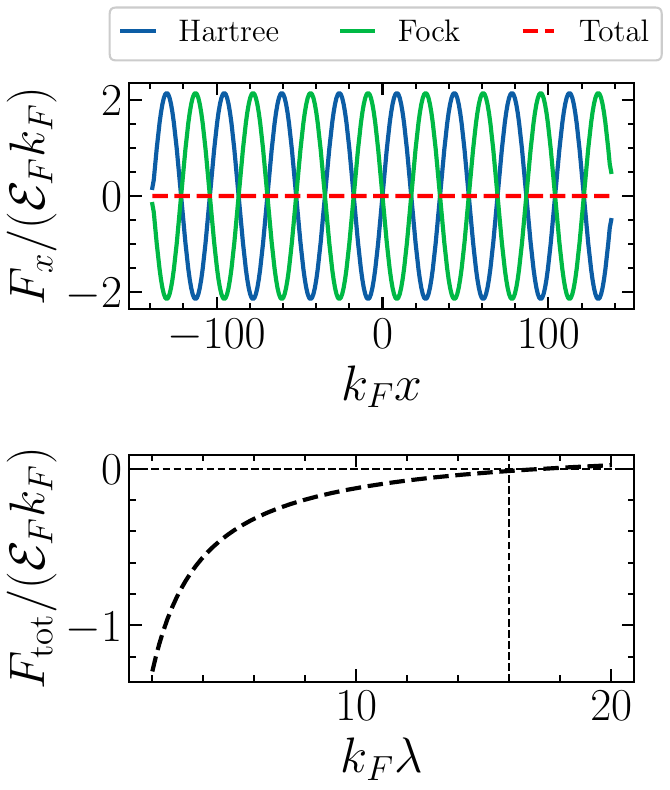}
    \caption{Top panel: Spatial profiles of the Hartree force $F^\text{H}(x)$, the effective Fock force $F^\text{F}(x)$, and their sum $F_{\mathrm{tot}}(x) = F^\text{H}(x) + F^\text{F}(x)$ for a perturbation wavelength in the intermediate regime $k_{F}\lambda \sim 1$ at $T/T_F \sim 1$. Bottom panel: Total force at a fixed position, $F_{\mathrm{tot}} = F^\text{H} + F^\text{F}$, as a function of the dimensionless modulation wavelength $k_{F}\lambda$. For long wavelengths $k_{F}\lambda \gg 1$ the total force follows the Hartree contribution. As the wavelength approaches the Fermi scale, $k_{F}\lambda \sim 1$, the exchange force grows and cancels the Hartree term at a critical wavelength where $F_{\mathrm{tot}}$ changes sign, marked by a vertical dashed line. For $k_{F}\lambda \lesssim 1$ the total force is dominated by exchange and points opposite to the electrostatic force.}
\label{fig_combined_force}
\end{figure}

To illustrate more directly how exchange modifies the local dynamics, we next inspect the total force
\begin{equation}
F_{\mathrm{tot}}(x) = F^\text{H}(x) + F^\text{F}(x)
\end{equation}
for a fixed temperature and density while varying the modulation wavelength. In the top panel of Fig.~\ref{fig_combined_force} we plot $F^\text{H}(x)$, $F^\text{F}(x)$ and their sum for a parameter choice inside the exchange--dominated region of Fig.~\ref{fig_HF_ratio}. Clearly, the Fock force is almost exactly out of phase with the Hartree force, and, as a consequence, the total force can be strongly reduced and even nearly vanish over an extended spatial region. This cancellation is a purely quantum effect, and in the classical limit the force always points along the electrostatic gradient.

The dependence of this cancellation on the modulation wavelength is summarized in the bottom panel of Fig.~\ref{fig_combined_force} depicting the total force at a fixed position as a function of $k_{F}\lambda$. For long wavelengths $k_{F}\lambda \gg 1$ we recover the classical behavior, with $F_{\mathrm{tot}}$ following the Hartree force. As $\lambda$ is reduced towards the Fermi scale, the magnitude of the exchange force grows until it compensates the Hartree contribution at a critical wavelength $k_{F}\lambda_{c} \sim 1$ where $F_{\mathrm{tot}}$ changes sign. For $k_{F}\lambda < 1$ the total force is strongly modified by exchange such that it points opposite to the electrostatic force. This sign reversal shows that, in the strongly quantum regime, exchange does not merely renormalize the strength of the restoring force but can qualitatively change the direction of the effective acceleration experienced by electrons.

From a microscopic point of view, the fact that $F^\text{F}$ opposes
the Hartree force can be traced back to the exclusion principle. The
antisymmetry of the many-body wave function generates an exchange hole
and a negative exchange contribution to the total energy, which
partially cancels the kinetic pressure and reduces the local
inverse compressibility of the electron gas. In the strongly quantum
regime this softening can overshoot the classical electrostatic response,
leading to an overscreening of the Hartree field and to the sign reversal
of $F_{\mathrm{tot}}$ observed in Fig.~\ref{fig_combined_force}.

\subsection{Exchange-mediated dynamical screening}
\label{subsec:dyn_screening}

The developed kinetic framework can also be used to study how exchange modifies the dynamical screening of a localized impurity in a degenerate 2DEG. This problem is directly relevant for semiconductor heterostructures~\cite{nixon1990potential}, where remote dopants, interface roughness and narrow gates generate inhomogeneous electrostatic landscapes that are screened by the mobile carriers~\cite{borisov2004dimensionality}. Most treatments of exchange in impurity screening are formulated in equilibrium or within linear response, where exchange is incorporated either via static energy functionals (renormalising the compressibility/Thomas--Fermi scale) or via local-field factors that dress the RPA susceptibility and screened interaction \cite{grinberg1987exchange,PhysRevB.74.155319,PhysRev.176.589}. These approaches are naturally expressed in $(q,\omega)$ space around an equilibrium state, so exchange enters as a renormalisation of response functions or mode dispersions. Here, instead, screening is obtained directly in real space and real time by evolving the Hartree--Fock--Wigner equation after a localised perturbation, giving the transient build-up of $\delta n(x,t)$ and the corresponding force profile without assuming an {\it a priori} dielectric function. 

The system is prepared at \(t=0\) in a spatially uniform Fermi--Dirac state of density \(n_0\) and temperature \(T\), and at the same time it is subjected to a static potential
\begin{equation}
U_{\mathrm{imp}}(x) = U_0 \exp\left[-\frac{(x-x_0)^2}{2\sigma_\text{imp}^2}\right],
\label{eq:Vimp_gauss}
\end{equation}
which models a charged impurity on the 2DEG. To characterize screening we monitor the density deviation
\begin{equation}
\delta n(x,t) = n(x,t) - n_0,
\end{equation}
together with the electrostatic force. We focus on the linear-response regime \(U_0 \ll \mathcal E_F\), where the density remains close to \(n_0\) and the impurity primarily excites wavevectors around \(k \sim 1/\sigma_\text{imp}\). The impurity width \(\sigma_\text{imp}\) is varied from the long-wavelength regime \(k_F \sigma_\text{imp} \gg 1\) down to the quantum regime \(k_F \sigma_\text{imp} \lesssim 1\).

\begin{figure}
    \centering
    \includegraphics[scale=0.47]{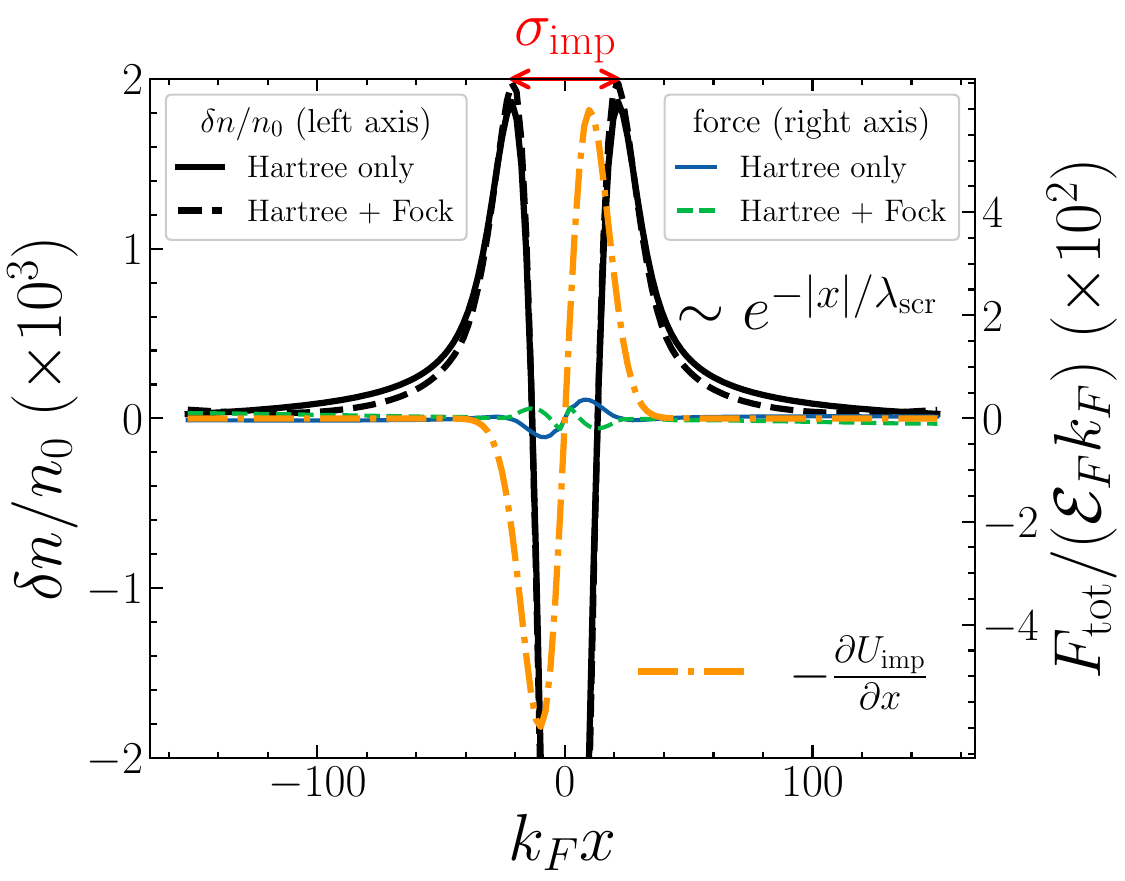}
    \caption{Left axis: Steady-state equilibrium density perturbation normalized to $n_0$ displaying Thomas-Fermi screening of a localized impurity with $k_F \sigma_\text{imp} = 10$ and $T = 5T_F$. Right axis: Total electrostatic force $F_{\mathrm{tot}}(x)$ vanishing far from the impurity. The plot shows the screened response obtained with Hartree theory alone and with the full Hartree--Fock interaction. Other parameters are \(r_s = 0.1\) and \(U_0 = 0.1 \mathcal E_F\).}
\label{fig_dyn_screening_maps}
\end{figure}

In the classical limit the impurity screening is fully described by the Thomas--Fermi theory. After a short transient the density approaches a stationary profile \(\delta n_{\mathrm{H}}(x)\), and the total electrostatic field
\begin{equation}
F_{\mathrm{tot}}(x) =
-\frac{\partial}{\partial x}\Phi^\text{H}(x) + F_{\mathrm{ext}}(x)
\end{equation}
vanishes way from the impurity, up to small plasma oscillations that decay by phase mixing. At high electron densities, typically $n \sim 10^{12}$--$10^{14}$ cm$^{-2}$ for the GaAs parameters considered here, and for impurities that vary slowly in space, $\sigma_\text{imp} \gg 1/k_{\mathrm{F}}$, the response of the 2DEG remains essentially linear. In this regime the induced density $\delta n(x)$ stays small compared to the background $n$, and the Hartree approximation already captures the screening quantitatively. Exchange acts only as a weak correction on top of the classical Thomas--Fermi response, while the bare impurity potential is almost completely flattened by the electron response, as shown in Fig.~\ref{fig_dyn_screening_maps}. Moreover, both $\delta n$ curves are well approximated by a decaying exponential outside the impurity volume. Hartree--Fock simulations differ only slightly from the Hartree ones, as including exchange leads to a modest enhancement of the screening wave vector and thus a slightly shorter screening length, in line with standard Thomas--Fermi theory~\cite{ashcroft1976solid,sham1966one}.

\begin{figure}
    \centering
    \includegraphics[scale=0.55]{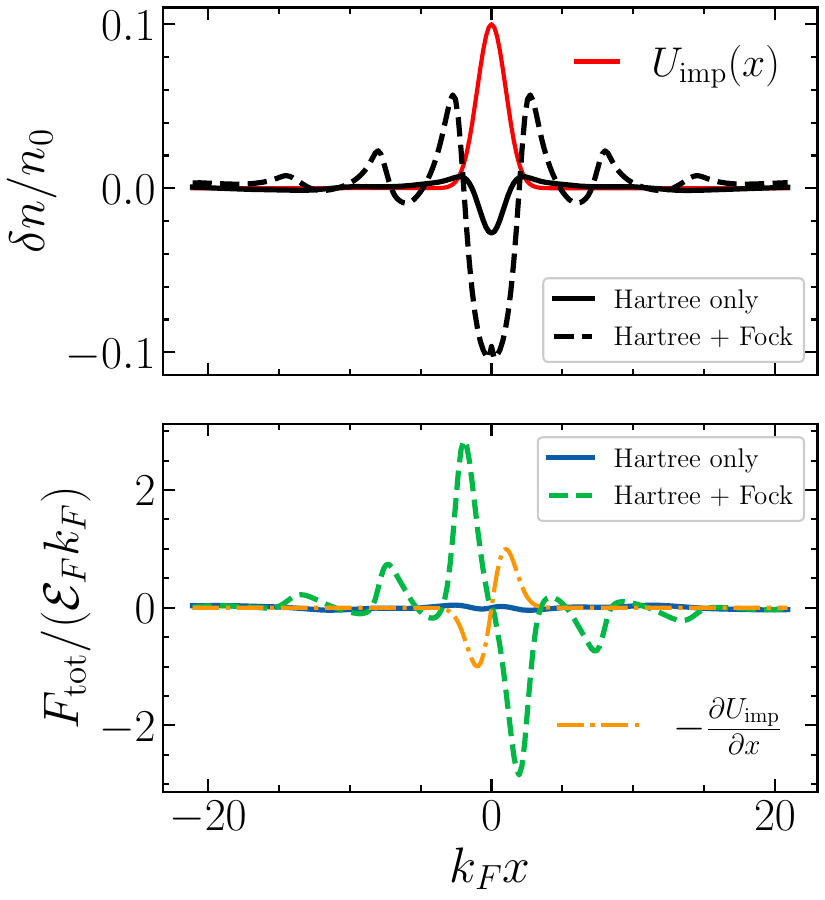}
    \caption{Top panel: Steady-state density perturbation normalized to $n_0$ displaying exchange-mediated screening of a localized impurity with $k_F \sigma_\text{imp}= 0.1$ and $T = 0.1T_F$, together with the external potential. Bottom panel: steady-state electrostatic profiles. Other parameters are \(r_s = 5\) and \(U_0 = \mathcal E_F\). }
\label{fig_static_profiles}
\end{figure}

When the electron density is lowered and the impurity becomes strongly localized, the screening response changes qualitatively and exchange effects play a dominant role. Two ingredients are crucial here. First, at low density the 2DEG becomes more strongly interacting as $r_s$ grows and the Fermi energy decreases, leading to an exchange energy per particle corresponding to a significant fraction of the kinetic energy. Second, a narrow impurity potential contains large Fourier components $q \sim 1/\sigma_\text{imp}$. When $q$ approaches or exceeds $2k_F$ the screening excites Friedel-like density oscillations with period $~\pi/k_F$, as shown in the top panel of Fig.~\ref{fig_static_profiles}. Contrarily, the classical density depletion $\delta n(x)$ monotonically decays outside the impurity ($|x|>\sigma_\text{imp}$). This is an example of a behavior that cannot be captured by a purely Hartree potential. In particular, for densities $n \lesssim 10^{12}$ cm$^{-2}$ (corresponding to $r_s$ of order unity or larger in GaAs) and narrow Gaussian impurities, we observe electronic overscreening, which manifests by an induced charge around the positively charged impurity that overcomes charge neutralization by producing a net negative charge in a region surrounding the impurity. According to the bottom panel of Fig.~\ref{fig_static_profiles}, including exchange effects modifies the otherwise vanishingly small electrostatic field, which becomes strongly attractive near the impurity, indicating a pronounced overcompensation of the test charge by the electronic cloud. This is reminiscent to the negative compressibility of the dilute 2DEG~\cite{eisenstein1994compressibility,schakel2001ground}, which states that it is energetically favorable to accumulate electrons in an already dense region. In Fig.~\ref{fig_static_profiles} this leads to a pronounced peak in the induced density $\delta n(x)$ at $x=0$ that exceeds the Hartree prediction by a large factor, indicating a much stronger local binding.

\begin{figure}
    \centering
    \includegraphics[scale=0.35]{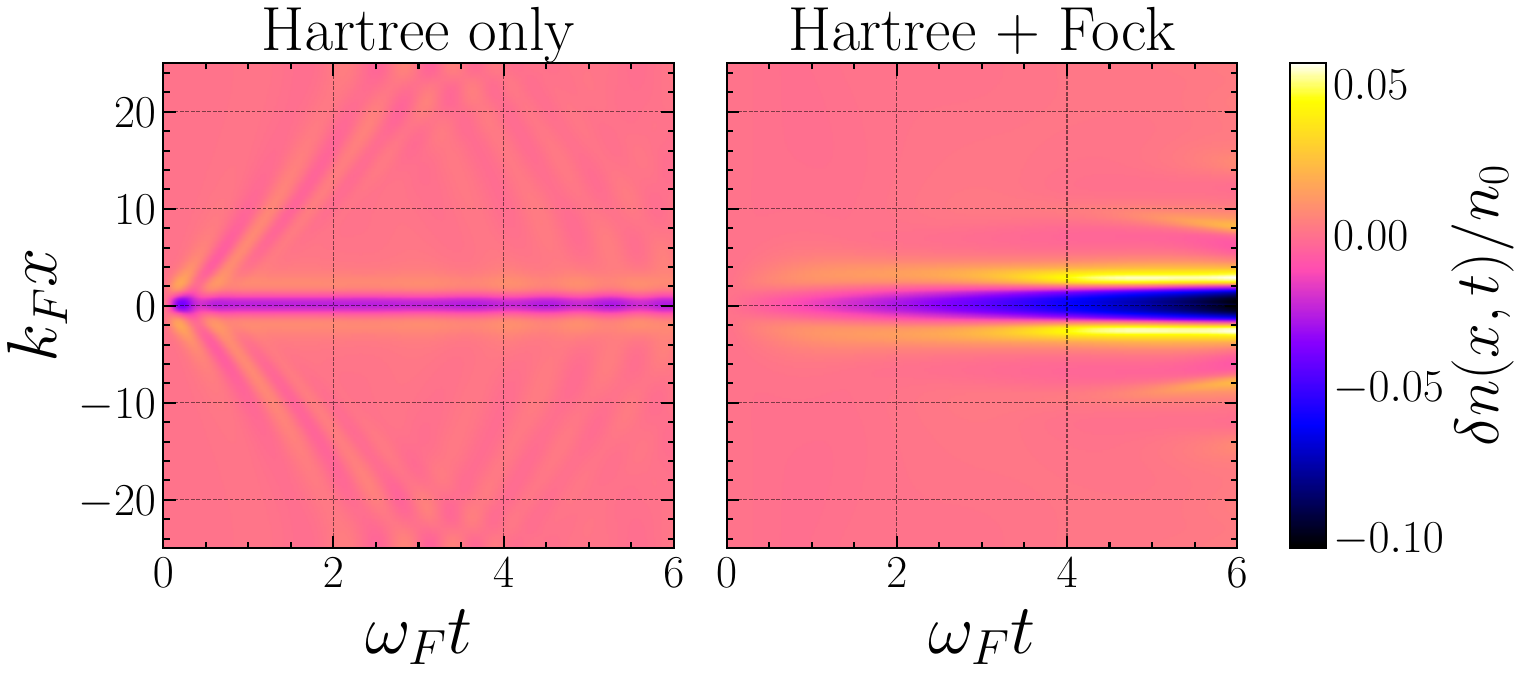}
    \caption{Space-time evolution of the density perturbation $\delta n(x,t)/n_0$ in the Hartree (left panel) and Hartree--Fock (right panel) cases. The simulation parameters are those of Fig.~\ref{fig_static_profiles}. For the classical evolution, density perturbations disperse away from the impurity, while in the Hartree--Fock case the density remains localized and deepens over time due to exchange effects. At the final times, Friedel-like oscillations are visible around the impurity in the Hartree--Fock case, due to the excitation of small wavelengths close of the order of $k_F^{-1}$.}
\label{fig_time_space_map}
\end{figure}

The space--time evolution of $\delta n(x,t)$ with and without exchange is shown in Fig.~\ref{fig_time_space_map} and highlights the fundamentally different relaxation pathways of the two regimes. In the Hartree case, density perturbations generated at the impurity disperse symmetrically, forming weak, rapidly damped wavefronts that propagate away from $x=0$. This behaviour is clearly visible in the left panel of Fig.~\ref{fig_time_space_map}, where the perturbation decays by phase mixing and the system quickly approaches a nearly flat late-time profile. In contrast, including exchange qualitatively alters the relaxation: instead of dispersing, the excess density remains localized near the impurity and, according to Fig.~\ref{fig_time_space_map}, gradually deepens as the system evolves. The impurity thus becomes dynamically self-trapped by its own exchange-enhanced charge cloud.

In the weak-coupling, high-density limit, the screening length extracted from the spatial decay of $\delta n(x)$ matches the Thomas--Fermi prediction $\lambda_{\mathrm{TF}}^{-1} = \sqrt{2\pi e^2 \nu(\mathcal E_F)/\varepsilon}$, where $\nu(\mathcal E_F)$ is the density of states at the Fermi level. As illustrated in Fig.~\ref{fig_dyn_screening_maps} (left axis), both Hartree and Hartree--Fock solutions exhibit exponential decay $\delta n \sim e^{-|x|/\lambda_{\mathrm{scr}}}$ with nearly identical $\lambda_{\mathrm{scr}}$ when $r_s \ll 1$ and $k_F \sigma_{\mathrm{imp}} \gg 1$. As the density is lowered or the impurity is sharpened, this correspondence breaks down. The emergence of Friedel-like oscillations and overscreening (Fig.~\ref{fig_static_profiles}, top panel) invalidates any single-length-scale description. The envelope of $\delta n(x)$ still decays, but with a substantially shorter effective screening length due to the increased exchange energy at low densities. In this regime, the usual notion of a monotonic screening length becomes ill-defined: instead, the impurity is surrounded by a region of alternating positive and negative induced charge whose spatial extent is set by the Fermi wavelength rather than by $\lambda_{\mathrm{TF}}$. The deep attractive well in the total force profile in Fig.~\ref{fig_static_profiles} (bottom panel) demonstrates that the exchange field can locally dominate the restoring force and effectively pins the electronic density.

\subsection{Exchange-driven antisymmetric instability in a coupled bilayer}

Having addressed the single-layer configuration, we now 
consider a richer geometry consisting of two parallel two-dimensional
electron gases separated by a distance $d$ and coupled through the
Coulomb interaction as described in Section~\ref{sec_four}. This configuration mimics double quantum wells and
dual-gated 2D devices, where negative
compressibility effects~\cite{eisenstein1992negative} and quantum capacitance~\cite{xia2009measurement} have been reported. Our aim is to show that, in
this regime, a kinetic description that neglects exchange misses an
entire class of dynamical instabilities, as anticipated in Section~\ref{sec_three}.

We model the two layers by distribution functions $f^{(1)}_{k}(x)$ and $f^{(2)}_{k}(x)$, each evolving in time under the Vlasov equation with both Hartree
and Fock mean fields. The coupling between the layers is promoted by long-range electrostatic forces according to Eq.~\eqref{Poisson_CD}. It is convenient to introduce symmetric and antisymmetric density
combinations,
\begin{equation}
    n_{\pm}(x,t) = n_{1}(x,t) \pm n_{2}(x,t),
\end{equation}
and similarly for their Fourier components $n_{\pm}(q,t)$. In the
Hartree sector the symmetric mode couples to the potential
$V_{+}(q) \sim (1 + e^{-|q|d})/|q|$, while the antisymmetric mode
couples to $V_{-}(q) \sim (1 - e^{-|q|d})/|q|$. For small $qd$ one
has $V_{-}(q) \simeq r_s d$, so that the restoring Hartree field
acting on $n_{-}$ is weak compared to the symmetric mode. Exchange, on
the other hand, acts independently in each sheet and can be included,
within a local approximation, as a negative contribution to the
compressibility. At long wavelengths, the linearized equation for the
antisymmetric mode can be derived from Eqs.~\eqref{dnDt} and~\eqref{djdt} as
\begin{equation}
    \frac{\partial^2}{\partial t^2} n_{-}(q,t)  + \Omega_{-}^{2}(q)\,n_{-}(q,t) = 0,
\end{equation}
with an effective frequency
\begin{equation}
    \Omega_{-}^{2}(q) \simeq \frac{q^{2}}{m} \Bigl[ \gamma_{-}^\text{H}(q) + \gamma_{-}^\text{F} \Bigr].
\end{equation}
Here $\gamma_{-}^\text{H}(q) = e^2 V_{-}(q)$ is the Hartree contribution and $\gamma^\text{F} = - \sqrt{2}e^2/(\pi^{3/2} \varepsilon \sqrt{n_0})$ is the (negative) exchange contribution to the inverse compressibility of the individual layers. When $|\gamma^\text{F}| > \gamma_{-}^\text{H}$ the effective
frequency becomes purely imaginary and the antisymmetric mode is
unstable. This indicates an exchange-driven instability of the bilayer
that has no analogue in a classical Vlasov description.

To test this prediction, we numerically solve the Hartree--Fock--Vlasov equation for two layers initialized with Fermi-Dirac momentum
distributions with different densities $n_{1} > n_{2}$ and equal temperature $T  \ll T_{\rm F,1},T_{\rm F,2}$, $T_{{\rm F},j}$ being the Fermi temperature of the $j-$th layer. Along the $x$ direction we impose a small antisymmetric modulation at a single wavevector $q_{m}$ as initial condition,
\begin{align} 
    n_{1}(x,0) &= n_{1}^{(0)} \bigl[1 + \Delta\cos(q_{m}x)\bigr],\\
n_{2}(x,0) &= n_{2}^{(0)} \bigl[1 -  \Delta\cos(q_{m}x)\bigr],
\end{align}
with $\Delta \ll 1$, so that only the antisymmetric mode $n_{-}$ is seeded.
To highlight the effect of exchange, we then evolve the system in time twice, once retaining only the Hartree field and once including both Hartree and Fock contributions. 
\begin{figure}
    \centering
    \includegraphics[scale=0.65]{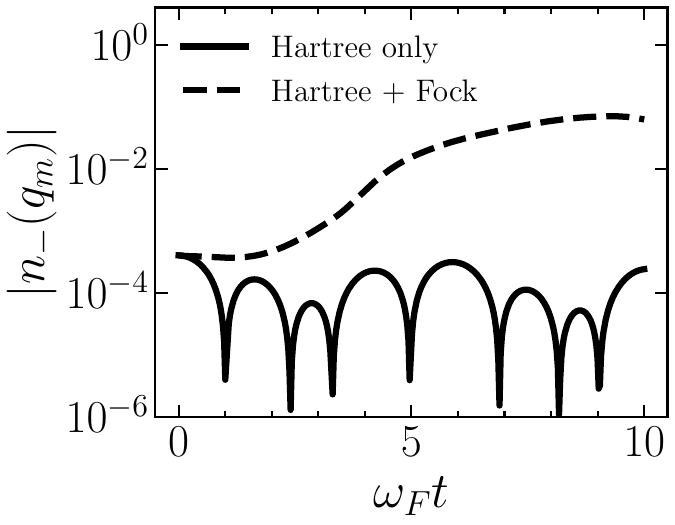}
    \caption{Time evolution of the antisymmetric Fourier component
$\lvert n_{-}(q_{m},t)\rvert$ for the coupled bilayer.
The blue curve shows the Hartree--only simulation, where the mode
performs small oscillations without secular growth.
The green curve shows the full Hartree--Fock dynamics, in which the
same mode grows by several orders of magnitude, signalling an
exchange--driven instability of the antisymmetric density channel.}
\label{fig_QuantumInst1_2}
\end{figure}

Figure~\ref{fig_QuantumInst1_2} shows the time evolution of the
antisymmetric Fourier amplitude $\lvert n_{-}(q_{m},t)\rvert$ for both simulations using $n_1^{(0)} = 5 \times 10^{10} \text{ cm}^{-2}$, $n_2^{(0)} = n_1^{(0)}/10$, $T = 5\,$K, $\Delta = 10^{-3}$ and $d = 28\,$nm. In the Hartree-only case the mode undergoes small
oscillations without secular growth, consistent with a stable plasma
oscillation whose frequency is set by $V_{-}(q_{m})$.
In contrast, when exchange is included, the same mode grows by several
orders of magnitude over the simulated time window, in line with the
expectation $\Omega_{-}^{2}(q_{m}) < 0$. The early-time growth is
approximately exponential, indicating a genuine linear instability.

\begin{figure}
    \centering
    \includegraphics[scale=0.6]{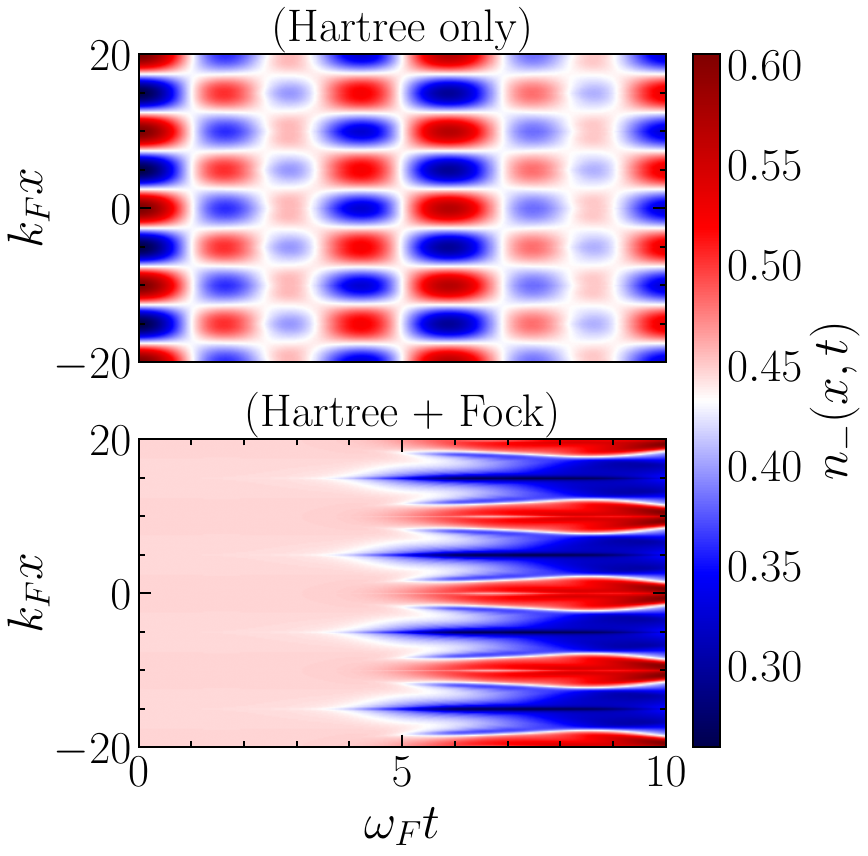}
    \caption{Spatiotemporal evolution of the antisymmetric density
$n_{-}(x,t) = n_{1}(x,t) - n_{2}(x,t)$ in the coupled bilayer.
Top panel: Hartree--only dynamics, where the initial antisymmetric
modulation remains small and simply oscillates in time.
Bottom panel: Hartree plus Fock dynamics, where the same initial
perturbation is exponentially amplified and evolves into a pronounced
stripe pattern of charge imbalance between the two layers.
The color scale is the same in both panels.}
\label{fig_twoSheet_forces}
\end{figure}

The corresponding real-space evolution of the antisymmetric density
$n_{-}(x,t)$ is displayed in Fig.~\ref{fig_twoSheet_forces}. In the
Hartree-only run, the initial modulation simply oscillates and spreads
over the domain, and its amplitude remains small.
When exchange effects are included, the antisymmetric mode grows coherently at the selected
wavevector, producing a pronounced pattern of charge imbalance between
the two layers. The maxima of $n_{-}$ sharpen in time and develop into a
quasi-stationary stripe pattern, signalling the onset of a nonlinear
charge-separation state driven by exchange.
At early times the Hartree
and Fock forces are comparable in magnitude and often opposite in sign,
so that the total force acting on the antisymmetric perturbation is
strongly reduced. This is the dynamical signature
of the overscreening identified in Sec.~\ref{sec_local_exchange} where the exchange field tends to invert the sign of the restoring force associated with $V_{-}(q)$, and drives the growth of $n_{-}$ when the effective compressibility becomes negative.

\begin{figure}
    \centering
    \hspace{-0.5cm}
    \includegraphics[scale=0.67]{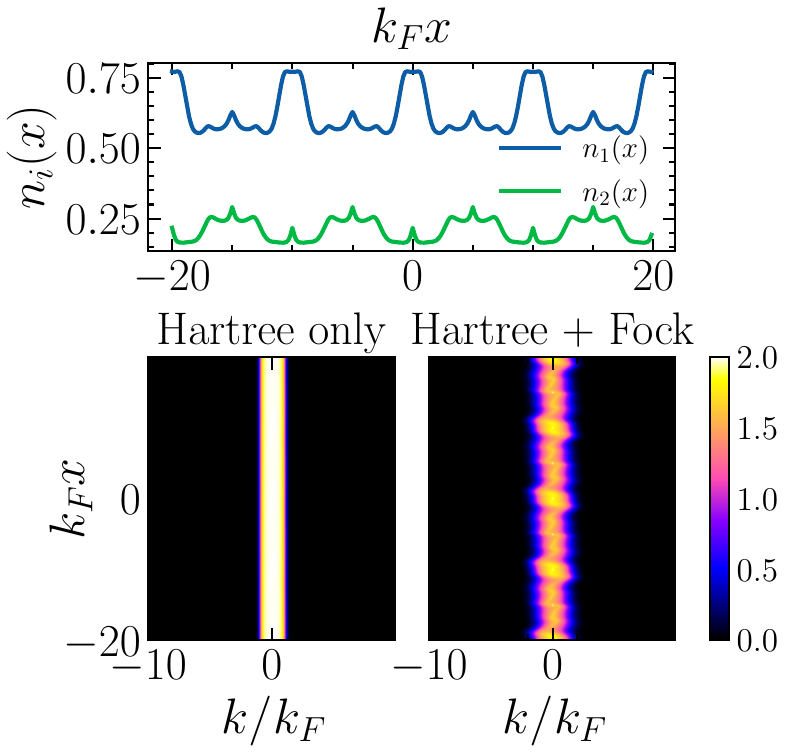}
    \caption{Top panel: Density profiles of the two layers in the Hartree--Fock simulation after saturation, showing strong spatial modulation and clear
out--of--phase density peaks and dips in the two layers.
The late--time profiles confirm that the instability selectively
amplifies the antisymmetric mode and produces a robust pattern of charge
separation between the sheets. Bottom panel: steady-state distribution function $f^{(1)}_{k}(x)$ in phase space for both Hartree and Hartree-Fock simulations. In the Hartree case the distribution remains close to its initial value, while in the Hartree-Fock case the distribution is strongly distorted due to the exchange-driven instability.}
\label{fig_twoSheet_profiles}
\end{figure}

Finally, Fig.~\ref{fig_twoSheet_profiles} shows the spatial density
profiles of the two layers at the final simulation time and including exchange. 
At later times, the dilute sheet becomes strongly modulated and develops
deep density dips at the positions where the dense layer exhibits local
maxima. The two profiles are dynamically out of phase, confirming that the
instability selectively amplifies the antisymmetric mode. In the
Hartree-only simulation the corresponding profiles remain close to their
initial values, and no comparable charge-separation pattern is observed.
The bottom panel of Fig.~\ref{fig_twoSheet_profiles} confirms these conclusions by showing the steady-state distribution function $f^{(1)}_{k}(x)$ in phase space. The Hartree--Fock panel reveals strong spatial modulations and a broadening in momentum, reflecting the onset of an exchange-driven instability. The phase-space distribution becomes corrugated, indicating that fermions experience momentum-dependent exchange forces that dynamically break translational symmetry and redistribute the density. These features are consistent with a spontaneous modulation of the Fermi sea driven by nonlocal exchange effects.

\subsection{Exchange corrections to Coulomb drag}

The kinetic formulation developed above can also be applied to nonequilibrium momentum transfer between spatially separated electron layers. In double quantum wells, Coulomb drag arises when a current driven in one (active) layer induces a frictional voltage in the other (passive) layer through interlayer electron--electron interactions. The effect was first observed in GaAs double quantum wells by Gramila \emph{et al.}~\cite{gramila1991mutual} and soon became a standard probe of electron--electron scattering in low-dimensional systems. On the theory side, early work by Jauho and Smith~\cite{jauho1993coulomb} and by Zheng and MacDonald~\cite{zheng1993coulomb} established that, in the weakly interacting Fermi-liquid regime, the drag resistivity $\rho_{\mathrm{D}}$ can be expressed in terms of the product of the dynamic structure factors of the two layers, leading to the well-known low-temperature scaling $\rho_{\mathrm{D}} \propto T^2$. Comprehensive reviews of Coulomb drag in 2D and more exotic systems are given in Refs.~\cite{asgari2019many,chen2015boltzmann}.

Exchange and correlation effects on drag have been considered mostly within linear-response and static-screening approximations. In ordinary charge drag, these corrections are commonly incorporated through local-field factors in the intralayer density response, which renormalise both the plasmon spectrum and the effective interlayer interaction~\cite{zheng1993coulomb}. In the related problem of spin Coulomb drag, D'Amico and Vignale showed that exchange and correlations can strongly suppress the spin diffusion constant by enhancing the friction between up- and down-spin currents~\cite{d2003spin,d2000theory}. More recent work has included exchange--correlation effects beyond the random-phase approximation, and has explored their impact on spin drag in low-dimensional systems~\cite{badalyan2008exchange}. In all these approaches, however, the electronic response is treated close to equilibrium and the exchange field enters not as a dynamical variable, but merely as a static renormalisation of response functions.

Our kinetic framework allows us to go beyond these limitations by computing the drag dynamically and self-consistently, with exchange incorporated at the level of the evolving distribution functions. In what follows we consider again the parallel-sheet configuration and assume that interlayer tunnelling is negligible. A drive current is applied to layer~1, and we monitor the induced momentum flow in layer~2 through $j_2(x) = \int \frac{d\mathbf k}{(2\pi)^2} \frac{\hbar k_x}{m} f^{(2)}_{k_x}(x)$. This approach captures not only the conventional Hartree-mediated drag but also the nonlocal Fock corrections that arise from correlations between identical fermions within each layer. As we show below, these exchange effects modify the drag in two distinct ways, either by renormalizing the intralayer response function and by reshaping the nonequilibrium distribution in the driven layer.

\subsubsection{Drag force from the kinetic equation}

To formulate Coulomb drag in coupled layers, we generalise the exchange-corrected Vlasov equation introduced above by including both an external drive and interlayer collision processes. The Wigner function $f^{(\ell)}_k(x,t)$ for layer $\ell=1,2$ obeys
\begin{equation}
\frac{\partial f^{(\ell)}}{\partial t}
+V(x,k,t)\frac{\partial f^{(\ell)}}{\partial x}
+ F_{\ell}(x,t) \frac{\partial f^{(\ell)}}{\partial k}
= \sum_{\ell '} \mathcal C_{\ell, \ell'}[f^{(\ell)}, f^{(\ell')}],
\label{eq:vlasov_bilayer}
\end{equation}
where the left-hand side is that of Eq.~\eqref{KinEq2}
to which we add an external drive $F_{\ell}^{\mathrm{ext}} = -e E_\ell^\text{ext}$. Moreover, for sufficiently small $d$ (typically of order $\sim 1/k_F$ or smaller), interlayer Coulomb collisions $\mathcal C_{\ell, \ell'} = \delta_{\ell',\overline{\ell}} \, \mathcal I_{\ell, \overline{\ell}}$ are not negligible and are known to contribute to the drag signal substantially~\cite{chen2015boltzmann}. The form of the interlayer collision integral for Coulomb-drag configurations has been derived in Ref.~\cite{messica2024hall} in the Born approximation, and reduces to a frictional force of the form
\begin{equation}
    \mathcal I_{\ell, \overline{\ell}}(x,k,t) = \tau^{-1} \Big[ u_{\overline{\ell}}(x,t) - u_\ell(x,t) \Big] \frac{\partial f^{(\ell)}}{\partial k},
\end{equation}
where $u_\ell(x,t) = \frac{1}{n_\ell(x,t)} \int \frac{d\mathbf k}{(2\pi)^2} \frac{\hbar \mathbf{k}}{m} f^{(\ell)}_{k}(x,t)$ is the local drift velocity in layer $\ell$ and $\tau^{-1}_\ell\equiv \tau^{-1}_\ell(d,T)$ is an interlayer momentum-relaxation rate,
\begin{equation}
    \tau^{-1}_\ell(d,T) = \frac{\pi }{32}\frac{(k_BT)^2}{v_F^{\ell} v_F^{\overline{\ell}}} \frac{\eta(3)}{k_F^{\ell} k_F^{\overline{\ell}}d^4}.
\end{equation}

The drag force on layer~2 is defined as the rate of change of its momentum due to the interlayer interaction. Multiplying Eq.~\eqref{eq:vlasov_bilayer} by $k$ and integrating over $k$ and $x$ we obtain 
\begin{align}
&\frac{dP_{2}}{dt}
= \int dx\, n_{2}(x,t)\, \Big[ F_{2}^\text{H}(x,t)-eE_2^\text{ext} \Big] \nonumber \\
&+ \int dx \int \frac{d\mathbf k}{(2\pi)^2} \Big[ F^\text{F}_{2}(x,k,t) f^{(2)}_{k}(x,t) + k      \mathcal I_{2, 1}(x,k,t) \Big]. \label{eq_momentum_change}
\end{align}
From Eq.~\eqref{eq_momentum_change} we indentify the contributions to the momentum change arising from interlayer interactions to be
\begin{align}
\left.\frac{dP_{2}}{dt}\right|_{2 \rightarrow 1} &= -\int dx\, n_{2}(x,t) \frac{\partial \Phi^\text{H}_{2, 1}}{\partial x} \nonumber \\
&+ m \tau^{-1} \int dx\, n_{2}(x,t) u_{2}(x,t) u_{1}(x,t), \label{eq_drag_force_components}
\end{align}
where $ \Phi^\text{H}_{\ell, \ell'}$ is the Hartree potential in layer $\ell$ created by the density in layer $\ell'$. From conservation of the total momentum, we may write 
\begin{equation}
    \frac{dP_1}{dt} = - \frac{dP_2}{dt} \equiv - \int dx\, n_2(x,t)\, \Pi_{2\rightarrow 1}(x,t),
\label{eq_drag_force_def}
\end{equation}
where $\Pi_{\ell\leftarrow \ell'}$ is defined from Eq.~\eqref{eq_drag_force_components} and governs momentum exchange.

The drag resistivity is obtained from the inverse of the conductivity matrix relating spatially averaged currents $J_{\ell}$ with applied fields $E_{\ell}^\text{ext}$,
\begin{equation}
\begin{pmatrix}
J_1\\[2pt] J_2
\end{pmatrix}
=
\begin{pmatrix}
\sigma_{11} & \sigma_{21}\\
\sigma_{21} & \sigma_{11}
\end{pmatrix}
\begin{pmatrix}
E_1^\text{ext}\\[2pt] E_2^\text{ext}
\end{pmatrix}.
\end{equation}
Imposing open circuit conditions in the passive layer, $J_2=0$, and solving for $E_2^\text{ext}$ yields
\begin{equation}
\rho_{\mathrm{D}}
=
-\,\frac{E_2^\text{ext}}{J_1}.
\label{eq:rhoD_sigma_drag}
\end{equation}
The diagonal conductivity $\sigma_{11}$ and the transconductivity $\sigma_{21}$ are extracted directly from the long time averages of $J_1(t)$ and $J_2(t)$ in the simulations.

In experiments the drag resistivity is defined under open circuit conditions in the passive layer, $J_2=0$, and can be expressed in terms of the conductivity matrix after inversion, 
\begin{equation}
\rho_{\mathrm{D}}
=
-\,\frac{\sigma_{21}}{\sigma_{11}^2-\sigma_{21}^2}. \label{eq_rhoD_sigma}
\end{equation}
In our simulations we apply a field only to layer 1 and set $E_2^\text{ext}=0$, so that the steady state currents directly yield $\sigma_{11}=J_1/E_1^\text{ext}$ and $\sigma_{21}=J_2/E_1^\text{ext}$. The expression above is then used to infer the drag resistivity that would be measured in an open circuit configuration of layer 2.

In order to identify the microscopic origin of the exchange enhancement of $\rho_{\mathrm{D}}$, it is convenient to characterise separately how efficiently momentum injected into the active layer is transmitted to the passive layer and how efficiently the passive layer converts this momentum transfer into a drag current. We therefore introduce two key quantities that can be easily evaluated dynamically from the simulated distributions. The first quantity is a momentum transfer efficiency,
\begin{equation}
\eta_{\mathrm{drag}}(t)
=
\frac{ \int dx\,n_2(x,t)\,\Pi_{2\leftarrow 1}(x,t)}
     {\int dx\,n_1(x,t)\,F^{\mathrm{ext}}_1}, \label{gamma_drag_def}
\end{equation}
where, according to Eq.~\eqref{eq_drag_force_components}, $\Pi_{2\leftarrow 1}(x)=F^{\mathrm{H}}_{2,{\rm inter}}(x)+m\tau^{-1}(u_1-u_2)$ is the Hartree net interlayer force density acting on layer 2. This dimensionless coefficient measures how effectively the momentum supplied by the drive field is transmitted across the barrier. The second quantity is the odd component of the distribution in the passive layer,
\begin{equation}
f_2^{\mathrm{odd}}(x,k)
=
\frac{1}{2}\left[f^{(2)}_{k}(x)-f^{(2)}_{-k}(x)\right], 
\end{equation}
which controls the drag current since only antisymmetric distortions in momentum space contribute to $j_2(x)$. We define the total odd weight in layer 2 as
\begin{equation}
\overline{f_2^{\mathrm{odd}}}(t)
=
\int dx \int dk\,\bigl|f_2^{\mathrm{odd}}(x,k,t)\bigr|. \label{eq_Wodd_def}
\end{equation}
By construction, $\eta_{\mathrm{drag}}$ probes the efficiency of interlayer momentum transfer while $\overline{f_2^{\mathrm{odd}}}$ measures how much of this transferred momentum is stored in a current carrying distortion of $f_2$.
From Eq.~\eqref{eq_rhoD_sigma} we see that, although no interlayer exchange force is present, $\rho_{\mathrm{D}}$ is still affected by exchange in essentially two ways. First, it modifies the intralayer density response of each sheet, changing the amplitude and phase of the density fluctuations that appear in $n_1$ and thus in $\Pi_{2\rightarrow 1}$. Second, the nonlocal exchange field reshapes the nonequilibrium distribution $f^{(1)}_{k}$, and consequently $u_1$, as well as the spectrum of excitations that participate in interlayer scattering beyond what is captured by a linearised picture.

\subsubsection{Numerical setup and exchange-induced corrections}

To quantify these effects we simulate two parallel 2DEGs with identical band parameters and densities, coupled through the interlayer Hartree kernel and interlayer frictional force. The initial condition consists of two uniform Fermi--Dirac distributions at density $n_0$ and temperature $T$. At $t=0$ we apply a constant in-plane electric field $E_1^\text{ext}$ to layer~1 by adding a uniform external force $F_1^{\mathrm{ext}} = -eE_1^\text{ext}$, while layer~2 is kept unbiased. For each set of parameters we evolve the coupled kinetic equations until the steady-state of Eq.~\eqref{eq_rhoD_sigma} is reached.

\begin{figure}
    \centering
    \hspace{-1cm}
    \includegraphics[scale=0.65]{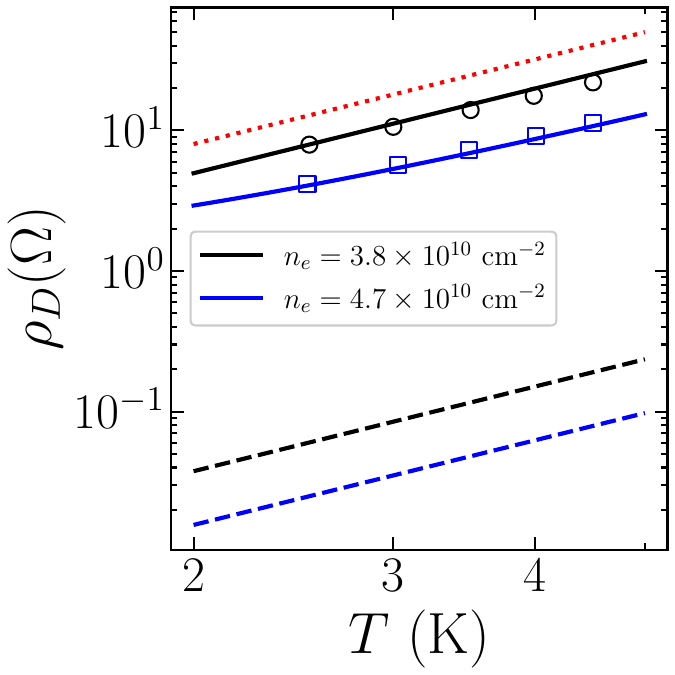}
\caption{
Drag resistivity $\rho_{\mathrm{D}}(T)$ obtained from the kinetic simulations for two electron densities, $n_e = 3.8 \times 10^{10}\,\mathrm{cm^{-2}}$ and $n_e = 4.7 \times 10^{10}\,\mathrm{cm^{-2}}$, and separation $d = 28\,\mathrm{nm}$. 
Solid lines include both Hartree and Fock contributions, while dashed lines show the corresponding classical result (exchange neglected). The data points are experimental measurements from Ref.~\cite{kellogg2002evidence} for the same simulated conditions, while 
the red dotted line is a reference $\rho_D \sim T^{2}$ line for comparison. 
}
\label{fig_drag_resistivity}
\end{figure}

\begin{figure}
    \centering
    \includegraphics[scale=0.7]{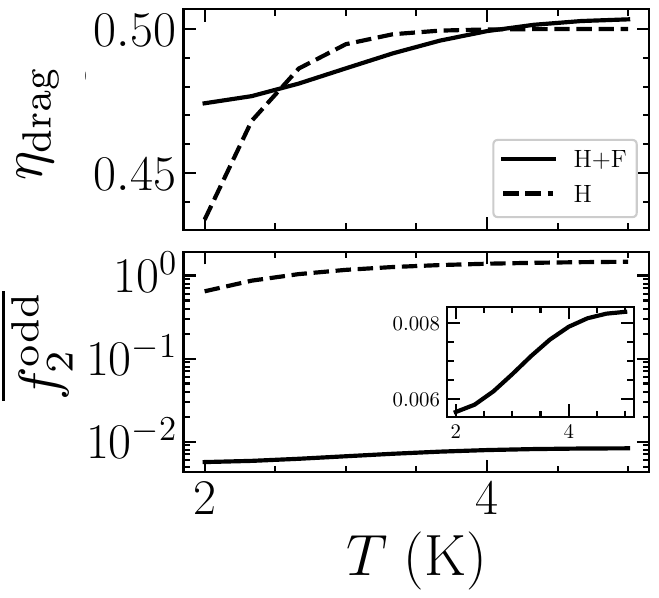}
    \caption{Temperature dependence of (top) the momentum-transfer efficiency 
$\eta_{\mathrm{drag}}$ and (bottom) the odd component 
$\overline{f^\text{odd}_2}$ of the passive-layer distribution for an electron 
density $n_e = 3.8 \times 10^{10}\,\mathrm{cm^{-2}}$ in steady state.  
Solid lines show the Hartree--Fock results, while dashed lines correspond 
to the Hartree-only simulation. The inset shows the variation of $\overline{f^\text{odd}_2}$ for the Hartree--Fock case on a linear scale.}
\label{fig_drag_Gamma_Wodd_vs_T}
\end{figure}

The temperature dependence of the drag resistivity obtained from the kinetic 
simulations is shown in Fig.~\ref{fig_drag_resistivity}. At low temperatures 
$T \lesssim 10\,\mathrm{K}$ all curves display the expected Fermi--liquid 
behavior $\rho_{\mathrm{D}} \propto T^{2}$, as indicated by the red dotted 
reference line. While the temperature exponent is unaffected by exchange, the 
magnitude of the drag resistivity is strongly enhanced when the Fock term is 
included. For both densities considered, $n_e = 3.8 \times 10^{10}\,\mathrm{cm^{-2}}$ and $4.7 \times 10^{10}\,\mathrm{cm^{-2}}$, the Hartree--Fock curves lie well 
above the Hartree-only results across the entire temperature range. The 
enhancement is particularly pronounced at the lower density, reflecting the 
increased nonlocality of the exchange field in dilute 2DEGs. Importantly, the 
magnitude and temperature dependence of $\rho_{\mathrm{D}}(T)$ obtained from the 
Hartree--Fock simulations are in good quantitative agreement with the 
measurements reported by Kellogg \textit{et al.}~\cite{kellogg2002evidence} (marked by data points in Fig.~\ref{fig_drag_resistivity}), 
whereas the Hartree-only results underestimate the drag by over an order of 
magnitude. To identify the microscopic origin of the enhanced drag, we examine the
quantities that mediate the relation between the applied drive in layer~1 and
the resulting response in layer~2. Figure~\ref{fig_drag_Gamma_Wodd_vs_T} shows
the temperature dependence of the interlayer momentum--transfer efficiency
$\eta_{\mathrm{drag}}$ (top) and of the odd component
$\overline{f^\text{odd}_2}$ of the passive--layer distribution (bottom).

According to Fig.~\ref{fig_drag_Gamma_Wodd_vs_T}, the
efficiency $\eta_{\mathrm{drag}}$ changes only weakly when exchange is included, as the Hartree and Hartree--Fock curves differ by less than
$\sim 10\%$, indicating that the total momentum transmitted across the barrier is nearly the same in both models. The enhancement of the drag resistivity therefore cannot be attributed to an increase in the net interlayer friction. The dominant effect arises instead from how exchange reshapes the \emph{response} of the passive layer to a given interlayer force. In previous sections we showed that in dilute 2DEGs the exchange force corrections often act in opposition to the Hartree force, effectively reducing the net acceleration experienced by
electrons. This mechanism was found to underlie overscreening, reversed effective acceleration, and the exchange-driven instability in the bilayer geometry. The same physics plays a crucial role here. When the interlayer force acts on layer~2, the resulting current-carrying distortion of its distribution is controlled by the \emph{net} force entering
the kinetic equation. Because the exchange field in layer~2 partially cancels the perturbing Hartree field generated by layer~1, the electrons in the passive layer accelerate less efficiently than they would under the Hartree force alone. As a result, for the same interlayer momentum transfer, the induced current in the passive layer is substantially smaller in the presence of exchange.

This reduced responsiveness is quantified by the odd component
$\overline{f^\text{odd}_2}$, which measures the part of $f^{(2)}_{k}(x)$ that contributes to the drag current. As seen in the lower panel of
Fig.~\ref{fig_drag_Gamma_Wodd_vs_T}, $\overline{f^\text{odd}_2}$ decreases by more than an order of magnitude when exchange is included. This does not imply that less momentum is transferred between layers, but rather reflects the fact that the passive layer requires a much larger distortion of its distribution to counteract the reduced effective acceleration caused by exchange term. Physically, the exchange field reshapes $f^{(1)}_{k}(x)$ and $f^{(2)}_{k}(x)$ by selectively suppressing excitations with the appropriate momentum asymmetry, thereby attenuating the odd part of the passive distribution needed to satisfy the steady-state condition.

The connection to the drag resistivity follows from the relation
$\rho_{\mathrm{D}} \approx -\sigma_{21}/\sigma_{11}^{2}$ valid when
$|\sigma_{21}| \ll |\sigma_{11}|$. Upon introducing exchange, the change in $\sigma_{11}$ is modest because the external field in layer~1 continuously injects momentum and offsets
much of the smoothing action of the exchange term. In contrast, $\sigma_{21}$
is directly connected to $\overline{f^\text{odd}_2}$, and therefore decreases
dramatically when exchange is included. Since the passive layer develops  much weaker odd distortion for the same interlayer force, a significantly lower compensating electric field must be generated to enforce the open-circuit condition $J_2 = 0$, leading to the enhanced drag resistivity observed in Fig.~\ref{fig_drag_resistivity}.

\section{Discussion and conclusions}\label{sec_conclusions}

This work established a quantum kinetic framework for two-dimensional electron gases in which exchange is treated at the Hartree--Fock level, revealing corrections to both the phase-space velocity and force stemming from a nonlocal exchange field. The resulting Hartree--Fock--Wigner equation provides a versatile platform to explore the dynamical consequences of exchange in low-dimensional conductors, capturing not only static renormalisations of compressibility and screening length, but also how exchange reshapes collective modes, promotes localisation, and alters nonequilibrium transport phenomena such as Coulomb drag. Our method goes beyond conventional approaches where exchange is introduced through static energy functionals or local-field factors by retaining the full phase-space dynamics of the Wigner function and the self-consistent evolution of the nonlocal Fock field.

From a practical standpoint, our results show that a mean-field description that includes exchange is already capable of accounting for several observations in dilute two-dimensional systems. Most notably, the Hartree--Fock calculations reproduce both the magnitude and temperature dependence of the enhanced Coulomb drag resistivity measured in low-density GaAs bilayers, in a parameter regime where standard Hartree or RPA-based theories fall short. Within our framework, this enhancement is traced to a redistribution of occupation in momentum space, as the exchange field partially cancels the interlayer Hartree force acting on the passive layer, thereby reducing its effective acceleration. As a result, a larger distortion of the passive-layer distribution is required to sustain a given drag current, leading to a substantial increase in the drag resistivity compared to classical result. 

The model developed here also opens several directions for future work in systems where exchange and hydrodynamics intertwine. In ultra-clean graphene and related materials, experiments have revealed viscous electron flow~\cite{bandurin2016negative}, Poiseuille-like transport~\cite{crossno2016observation}, and hydrodynamic signatures~\cite{lucas2016transport,lucas2018hydrodynamics} over a broad temperature window. Embedding our exchange-corrected kinetic equation within a hydrodynamic closure offers a route to quantify how nonlocal Fock forces modify effective viscosities, sound modes, and the onset of hydrodynamic instabilities in such electron fluids. This is particularly relevant in regimes where the electron-electron scattering rate is large enough to establish local equilibrium, but exchange remains comparable to the thermal scale and can still alter the collective flow.

Another natural application concerns plasmonics and nanophotonics in two-dimensional materials. Strongly confined plasmons and polaritons in graphene, semiconductor heterostructures, and van der Waals stacks are routinely operated in regimes of moderate carrier density and low temperature, where exchange and correlations are non-negligible~\cite{koppens2011graphene,fei2012gate,fei2012gate,basov2016polaritons}. Our kinetic framework provides the microscopic input needed to describe how exchange renormalises plasmon dispersion, damping, and gain in patterned or inhomogeneous structures, and can be extended to treat Moiré superlattices~\cite{du2024nonlinear}, twisted bilayers~\cite{behura2021moire}, or engineered one-dimensional channels~\cite{nikitin2011edge} where bound states and localised modes play a key role. In these contexts, the ability to follow the full phase-space dynamics in the presence of spatially structured potentials or gates is a clear advantage over purely linear-response or local-density approaches. Beyond charge transport, the same formalism can be generalised to include spin degrees of freedom, thereby enabling a unified description of spin Coulomb drag~\cite{d2003spin}, spin-charge coupled modes~\cite{weber2005observation}, and spin-dependent instabilities in low-dimensional conductors~\cite{parthenios2025spin}. Incorporating spin into the Wigner-function framework and extending the Hartree--Fock decoupling to spin-resolved interactions would make it possible to revisit long-standing open questions on spin transport and relaxation in 2DEGs from a fully kinetic perspective~\cite{wu2010spin,raichev2025momentum,belykh2020anomalous}.

\bibliography{bibtex.bib}

\end{document}